\documentclass[aps,prb,twocolumn,superscriptaddress]{revtex4}
\usepackage{graphicx}
\usepackage{amssymb}
\usepackage{amsmath}
\usepackage{epsfig}
\usepackage{array}
\usepackage{color}
\usepackage{multirow}
\usepackage{dsfont}
\usepackage{bm}
\usepackage{ulem} 

\begin{document}
\title{Multipole-fluctuation pairing mechanism of $d_{x^2-y^2}+ig$ superconductivity in Sr$_2$RuO$_4$}

\author{Yutao Sheng}
\affiliation{Beijing National Laboratory for Condensed Matter Physics and
Institute of Physics, Chinese Academy of Sciences, Beijing 100190, China}
\affiliation{School of Physical Sciences, University of Chinese Academy of Sciences, Beijing 100049, China}
\author{Yu Li}
\affiliation{Kavli Institute for Theoretical Sciences, University of Chinese Academy of Sciences, Beijing 100190, China}
\author{Yi-feng Yang}
\email[]{yifeng@iphy.ac.cn}
\affiliation{Beijing National Laboratory for Condensed Matter Physics and
Institute of Physics, Chinese Academy of Sciences, Beijing 100190, China}
\affiliation{School of Physical Sciences, University of Chinese Academy of Sciences, Beijing 100049, China}
\affiliation{Songshan Lake Materials Laboratory, Dongguan, Guangdong 523808, China}

\date{\today}

\begin{abstract}
 Despite of many experimental and theoretical efforts, the pairing symmetry of superconductivity in Sr$_2$RuO$_4$ remains undecided. The accidentally degenerate $d_{x^2-y^2}+ig$ is consistent with most current experiments and seems to be one of the most probable candidates, but we still lack a satisfactory theoretical mechanism for its appearance. Here we construct a phenomenological model combining realistic electronic band structures and all symmetry-allowed multipole fluctuations as potential pairing glues, and make a systematic survey of major pairing states within the Eliashberg framework. Our calculations show that $d_{x^2-y^2}+ig$ can arise naturally from the interplay of antiferromagnetic, ferromagnetic, and electric multipole fluctuations whose coexistence is manifested in previous experiments and calculations. Our work provides a physically reasonable basis supporting the possibility of $d_{x^2-y^2}+ig$ pairing in superconducting Sr$_2$RuO$_4$.
\end{abstract}

\maketitle

\section{Introduction}
For over two decades, superconductivity in Sr$_2$RuO$_4$ had been proposed to be of odd-parity spin-triplet pairing both in theory and in experiment \cite{Maeno1994,Rice1995,Ishida1998,Mackenzie2003,Mackenzie2017}. This belief was recently overturned when refined nuclear magnetic resonance (NMR) \cite{Pustogow2019,Ishida2020,Chronister2021} and polarized neutron scattering (PNS) \cite{Petsch2020} experiments detected a drop in the spin susceptibility below $T_c$.
Muon spin relaxation ($\mu$SR) and polar Kerr effect revealed time-reversal symmetry breaking (TRSB) of the superconducting order parameter \cite{Luke1998,Xia2006}.
A line-node gap was then supported by specific heat \cite{NishiZaki2000,Deguchi2004,Kittaka2018}, penetration depth \cite{Bonalde2000}, thermal conductivity \cite{Suzuki2002,Hassinger2017}, spin-lattice relaxation rate \cite{Ishida2000}, and quasiparticle interference imaging from scanning tunneling microscope (STM) \cite{Sharma2020}. Candidate proposals of two-component TRSB order parameters include $d_{x^2-y^2}+ig$ \cite{Kivelson2020,Clepkens2021}, $s+id_{x^2-y^2}$ \cite{Romer2019,Romer2020}, $s+id_{xy}$ \cite{Romer2021,Bhattacharyya2021}, chiral or helical or mixed $p$-wave \cite{Mackenzie2003,Scaffidi2014,Roising2019,Wang2019,Ramires2019,Scaffidi2020,Gupta2020,Ikegaya2020,Chen2020,Wang2020,Huang2021}, $d_{xz}+id_{yz}$ \cite{Zhang2021}, and exotic interorbital pairings \cite{Huang2019,Kaba2019,Gingras2019,Suh2020,Kaser2021,Zhang2020}.

Further constraints can be extracted from several latest experiments. A detailed NMR analysis reported an upper limit of the condensate magnetic response and excluded all purely odd-parity states (such as $p_x+ip_y$) \cite{Chronister2021}. STM measurement pointed towards a nodal direction along the zone diagonal, supporting a dominant $d_{x^2-y^2}$ component \cite{Sharma2020}. Although later analysis suggested that $s+id_{xy}$ with accidental nodes near the zone diagonal might also explain the STM data \cite{Bhattacharyya2021}, it was often considered to be incompatible with the electronic structure of Sr$_2$RuO$_4$ \cite{Ghosh2021}. In ultrasound experiments, a thermodynamic discontinuity was reported in the shear elastic modulus $c_{66}$ \cite{Benhabib2021,Ghosh2021}, which excluded $s+id_{x^2-y^2}$. $\mu$SR measurements reported the split (unsplit) of the superconducting transition under uniaxial (hydrostatic) pressure \cite{Grinenko2021a,Grinenko2021b} and supported the symmetry-protected $d_{xz}+id_{yz}$ pairing, but specific heat measurement fails to see the split under uniaxial pressure \cite{Li2021}. Moreover, $d_{xz}+id_{yz}$ normally requires a jump in the $(c_{11}-c_{22})/2$ modulus which was not observed in the ultrasound experiment. 

Thus, the accidentally degenerate $d_{x^2-y^2}+ig$ seems to be most probable one among all candidates \cite{Kivelson2020}. It agrees with most of the above experiments, although a modification based on strain inhomogeneity might be needed to explain the absence of an evident specific heat jump at the TRSB transition under uniaxial pressure \cite{Yuan2021,Willa2021}. However, the occurrence of $d_{x^2-y^2}+ig$ lacks a strong theoretical support so far. In particular, it is unclear how the $g$-wave can arise and become accidentally degenerate with the dominant $d_{x^2-y^2}$ component. A recent theory can indeed derive the $g$-wave but requires a sizeable momentum-dependent spin-orbit coupling (SOC) beyond the first-principles prediction \cite{Clepkens2021}. Hence, a fully consistent explanation of the pairing symmetry in Sr$_2$RuO$_4$ has not been achieved.

In this work, we explore the possibility of $d_{x^2-y^2}+ig$ by constructing a general model Hamiltonian that combines realistic band structures from angle-resolved photoemission spectroscopy (ARPES) and multipole pairing interactions allowed by symmetry for the spin-orbit coupled Ru-$4d$ electrons. The superconducting gap structures are then evaluated systematically by solving the linearized Eliashberg equations with antiferromagnetic (AFM), ferromagnetic (FM), electric multipole fluctuations and their mixtures. We find that the $d_{x^2-y^2}+ig$ (pseudospin) singlet pairing can actually be generated by the interplay of these three multipole pairing interactions within a reasonable parameter range. This provides a natural physical basis for the occurrence of $d_{x^2-y^2}+ig$. We will also discuss briefly the conditions for other pairing states within our theoretical framework.

\section{Model}
Crystal field splitting and SOC are considered of equal importance in Sr$_2$RuO$_4$ \cite{Veenstra2014,Zhang2016,Mackenzie2017}. To capture the pairing symmetry, it is convenient to construct a general model Hamiltonian based on the multipole representation of the pairing interactions. By Stevens operator-equivalent technique, the multipole operators $\hat Q^{jkq}$ ($k=0,1,\dots,2j$; $q=-k,-k+1,\dots,k$) for a given angular momentum $j$ can be obtained from the $(2j+1)\times(2j+1)$ tensor operator $\hat J_{kq}$ satisfying \cite{Stevens1952,Inui1990}:
\begin{equation}
\begin{aligned}
& \hat J_{kk} = (-1)^k\sqrt{\frac{(2k-1)!!}{(2k)!!}} (\hat J_+)^k,
\\
& [\hat J_{\pm},\hat J_{kq}] = \sqrt{(k\mp q)(k\pm q+1)} \hat J_{k,q\pm1}\ (q<k),
\end{aligned}
\end{equation}
where $\hat J_{\pm}$ is the raising/lowering operator within the corresponding $j$-subspace. These multipole operators are further projected into the irreducible representation (IR) $\Gamma$ of the $D_{4h}$ point group of Sr$_2$RuO$_4$ and denoted as $\hat Q^{j\Gamma\alpha}$ for the $\alpha$-th component in $\Gamma$  \cite{Kusunose2008,Watanabe2018}. Table \ref{tab1} gives all multipole operators for the $j=3/2$ and $5/2$ manifolds of Ru-$4d$ electrons according to their IRs and ranks. The electric multipoles are of even-rank and time-reversal symmetric and listed on the top of the table, while on the bottom are the magnetic multipoles (odd-rank and time-reversal antisymmetric) \cite{Ikeda2012,Watanabe2018}. More details on the definition of these multipole operators can be found in Appendix A.

\begin{table}
\caption{\label{tab1}Multipole operators classified according to the irreducible representations $\Gamma$ of $D_{4h}$ point group based on the operator-equivalent technique. The $j=5/2$ manifold contains operators from rank 0 to rank 5 (monopole $\mathds{1}$; dipole $J$; quadrupole $O$; octupole $T$; hexadecapole $H$; dotriacontapole $D$), while multipole operators in $j=3/2$ are up to rank 3 (monopole $\mathds{1}$; dipole $J$; quadrupole $O$; octupole $T$). The subscript $g$ marks inversion symmetric representations and the superscripts $+/-$ denote time-reversal symmetric/antisymmetric ones. The subscripts of multipole operators are related to the tesseral harmonics in $O_h$ group or cubic harmonics \cite{Ikeda2012,Kusunose2008,Lage1947}. For simplicity, we have used the same symbols for both $j$-spaces. More details are explained in Appendix A.}
\renewcommand\arraystretch{1.5}
\centering
 \setlength{\tabcolsep}{4.5mm}{
\begin{tabular}{ccc}
\hline
\hline 
IR ($\Gamma$) & $\hat Q^{j=3/2,\Gamma\alpha}$ & $\hat Q^{j=5/2,\Gamma\alpha}$ \\
\hline  
& \multicolumn{2}{c}{Electric multipole operators} \\
$A_{1g}^+$ & $\hat{\mathds{1}}$, $\hat O_{20}$ & $\hat{\mathds{1}}$, $\hat O_{20}$, $\hat H_0$, $\hat H_4$ 	\\
$A_{2g}^+$ & & $\hat H_{za}$ 	\\
$B_{1g}^+$ & $\hat O_{22}$ & $\hat O_{22}$, $\hat H_2$ \\
$B_{2g}^+$ & $\hat O_{xy}$ & $\hat O_{xy}$, $\hat H_{zb}$ \\
\multirow{2}{*}{$E_{g}^+$} & \multirow{2}{*}{$(\hat O_{xz},\hat O_{yz})$} & $(\hat O_{xz},\hat O_{yz})$, $(\hat H_{xa},\hat H_{ya})$, \\
& & $(\hat H_{xb},\hat H_{yb})$ \\
\hline
& \multicolumn{2}{c}{Magnetic multipole operators} \\
$A_{1g}^-$ & & $\hat D_4$ \\
$A_{2g}^-$ & $\hat J_z$, $\hat T_{za}$ & $\hat J_z$, $\hat T_{za}$, $\hat D_{za1}$, $\hat D_{za2}$ \\
$B_{1g}^-$ & $\hat T_{xyz}$ & $\hat T_{xyz}$, $\hat D_2$ \\
$B_{2g}^-$ & $\hat T_{zb}$ & $\hat T_{zb}$, $\hat D_{zb}$ \\
\multirow{3}{*}{$E_{g}^-$} & $(\hat J_x,\hat J_y)$, & $(\hat J_x,\hat J_y)$, $(\hat T_{xa},\hat T_{ya})$, \\
& $(\hat T_{xa},\hat T_{ya})$, & $(\hat T_{xb},\hat T_{yb})$, $(\hat D_{xa1},\hat D_{ya1})$, \\
& $(\hat T_{xb},\hat T_{yb})$ & $(\hat D_{xa2},\hat D_{ya2})$, $(\hat D_{xb},\hat D_{yb})$ \\
\hline
\hline
\end{tabular}
}
\end{table}

\begin{figure}[t]
\begin{center}
\includegraphics[width=8.6 cm]{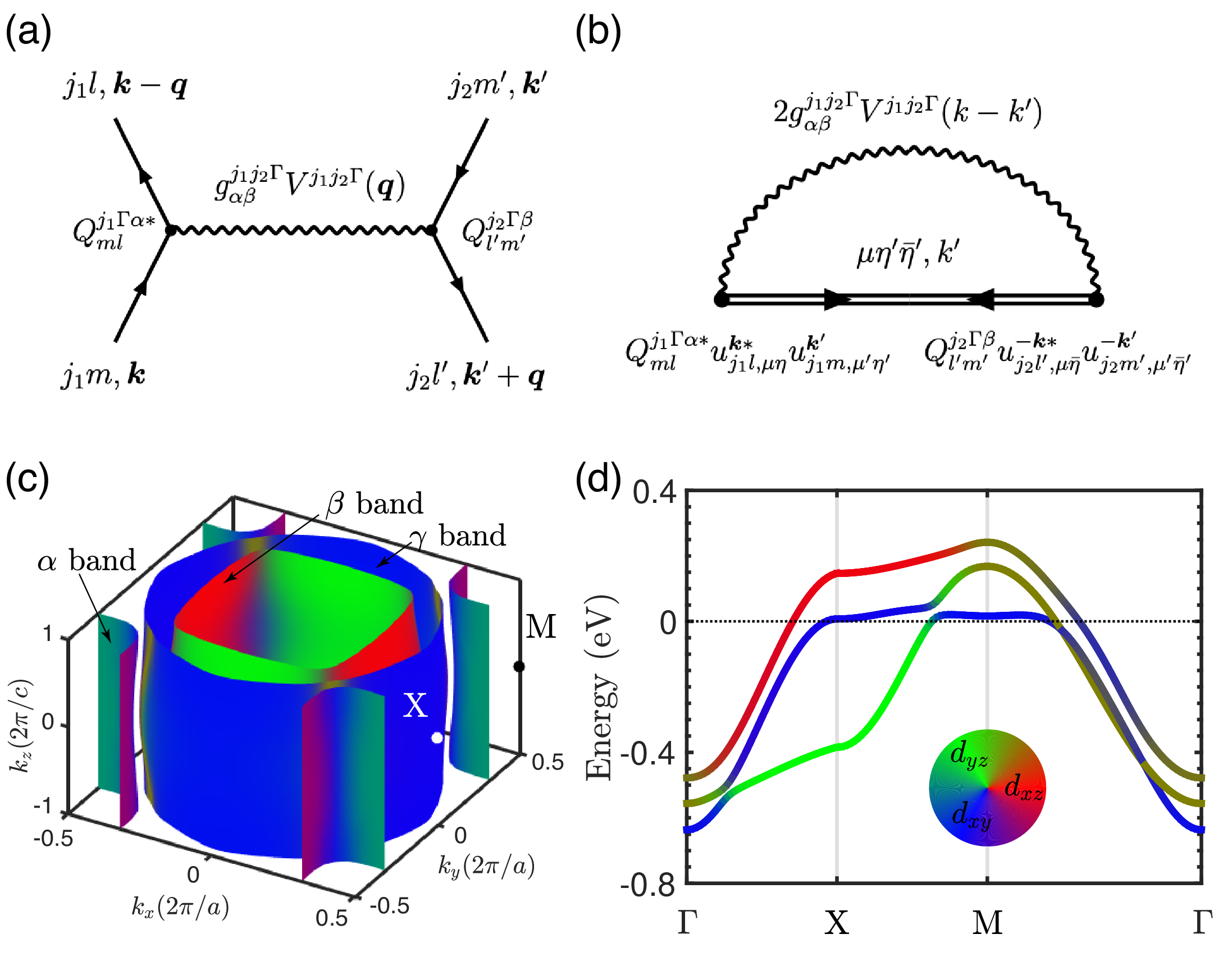}
\caption{(a) Illustration of the multipole interaction $\hat Q^{j_1\Gamma\alpha}\hat Q^{j_2\Gamma\beta}$. (b) The Feynman diagram of the anomalous self-energy $\psi_{\mu\eta\bar\eta}$ from multipole pairing interactions within the Eliashberg framework. We use $k=({\bf{k}},i\omega_n)$ for simplicity. (c) The 3D Fermi surfaces with $(d_{\text{xz}},d_{\text{yz}},d_{\text{xy}})$ orbital characters derived from the TB Hamiltonian $H_{\text{3D}}$. (d) Orbital-resolved band structures along a high-symmetry line on the $k_z=0$ plane of the Brillouin zone. The inset shows the colors for three orbitals.}
\label{fig1}
\end{center}
\end{figure}

We then write down a general interaction containing all symmetry-allowed multipole fluctuations as potential superconducting pairing glues:
\begin{equation}
\begin{aligned}
H_{\text{int}} = & - \sum_{j_1j_2}{\sum_{\Gamma\alpha\beta}}\sum_{{\bf{q}}} g^{j_1j_2\Gamma}_{\alpha\beta} V^{j_1j_2\Gamma}({\bf{q}}) \hat Q^{j_1\Gamma\alpha\,\dag}({\bf{q}}) \hat Q^{j_2\Gamma\beta}({\bf{q}})
\\
= & - \sum_{j_1j_2}{\sum_{\Gamma\alpha\beta}}\sum_{{\bf{q}},{\bf{k}},{\bf{k}}'} \sum_{lml'm'} g^{j_1j_2\Gamma}_{\alpha\beta} V^{j_1j_2\Gamma}({\bf{q}}) Q^{j_1\Gamma\alpha*}_{ml}Q^{j_2\Gamma\beta}_{l'm'} 
\\
& \times c^{\dag}_{j_1l,{\bf{k}}-{\bf{q}}} c_{j_1m,{\bf{k}}} c^{\dag}_{j_2l',{\bf{k}}'+{\bf{q}}} c_{j_2m',{\bf{k}}'},
\label{hamiltonian}
\end{aligned}
\end{equation}
where $\hat Q^{j\Gamma\alpha}({\bf{q}}) =\sum_{{\bf{k}},lm}Q^{j\Gamma\alpha}_{lm}c^{\dag}_{jl,{\bf{k}}+{\bf{q}}} c_{jm,{\bf{k}}}$ and $c_{jm,\bf{k}}$ ($c_{jm,\bf{k}}^\dag$) is the electron annihilation (creation) operator with $\bf{k}$ being the momentum and $m$ the $z$-projection of the total angular momentum $j$. The matrix elements $Q^{j\Gamma\alpha}_{lm}$ are normalized with $Q^{j\Gamma\alpha}_{lm} \rightarrow Q^{j\Gamma\alpha}_{lm} / \sqrt{\sum_{l'm'} |Q^{j\Gamma\alpha}_{l'm'}|^2}$ for comparison of different multipole fluctuations, $V^{j_1j_2\Gamma}(\bf{q})$ is the momentum-dependent interaction vertex, and $g^{j_1j_2\Gamma}_{\alpha\beta}$ controls the fluctuation strength between the multipole components $j_1\Gamma\alpha$ and $j_2\Gamma\beta$, as illustrated in Fig. \ref{fig1}(a). The values of $g^{j_1j_2\Gamma}_{\alpha\beta}$ are highly restricted as the multipole product should be projected to the identity representation of the crystallographic point group to keep the overall symmetry of the Hamiltonian. Thus only multipoles belonging to the same IR ($\Gamma$) can be coupled, but they could have different angular momentum ($j_1\not=j_2$) due to  comparable SOC and crystal field potential \cite{Khaliullin2013,Das2018}. For the two-dimensional IR $E^{\pm}_{g}$, such projection yields $(\hat Q^{j_1\Gamma\alpha}_{x}\hat Q^{j_2\Gamma\beta}_{x}+\hat Q^{j_1\Gamma\alpha}_{y}\hat Q^{j_2\Gamma\beta}_{y})/2$, which will be denoted as $\hat Q^{j_1\Gamma\alpha}_{r}\hat Q^{j_2\Gamma\beta}_{r}$ for simplicity. There are a total number of 6 electric multipole fluctuation channels and 11 magnetic multipole fluctuation channels in the $j=3/2$ manifold (listed at the bottom of Fig.~\ref{fig2}), 23 electric components and 38 magnetic components in the $j=5/2$ manifold (bottom of Fig.~\ref{fig7}), 15 electric and 30 magnetic $j$-mixed ($j_1\not=j_2$) multipole channels (bottom of Fig.~\ref{fig8}) that are allowed by symmetry in Sr$_2$RuO$_4$. For example, using the electric multipole operators listed in Table \ref{tab1}, we can generate 6 electric multipole fluctuation channels for $j=3/2$: $\hat{\mathds{1}}\hat{\mathds{1}}$, $\hat{\mathds{1}}\hat O_{20}$, $\hat O_{20}\hat O_{20}$, $\hat O_{22}\hat O_{22}$, $\hat O_{xy}\hat O_{xy}$ and $\hat O_{rz}\hat O_{rz}$, which are listed at the bottom of Fig.~\ref{fig2} according to the IRs and ranks of the corresponding multipole operators.

The above procedures lay out a general phenomenological framework for studying electron pairing induced by multipole fluctuations. To apply it to Sr$_2$RuO$_4$, we consider the following three dimensional (3D) tight-binding (TB) model, $H_{\text{3D}}=H_{\text{2D}}+H_z$, where $H_{\text{2D}}=\sum_{{\bf{k}},s} \psi^\dag_{s}({\bf{k}})h_0({\bf{k}},s)\psi_{s}({\bf{k}})$ describes the $k_z$-independent band structure from ARPES measurements \cite{Zabolotnyy2013}. $\psi_{s}({\bf{k}})=[c_{\text{xz},s}({\bf{k}}),c_{\text{yz},s}({\bf{k}}),c_{\text{xy},-s}({\bf{k}})]^T$ is the basis of  the low-lying Ru-$4d$ $t_{2g}$ orbitals $(d_{\text{xz}},d_{\text{yz}},d_{\text{xy}})$. We have
\begin{equation}
h_0({\bf{k}},s) = 
\begin{pmatrix}
\epsilon^{\text{xz}}_{{\bf{k}}}-\mu_0 & \epsilon^{\text{off}}_{{\bf{k}}}-is\lambda_{\text{SOC}} & i\lambda_{\text{SOC}} \\
\epsilon^{\text{off}}_{{\bf{k}}}+is\lambda_{\text{SOC}} & \epsilon^{\text{yz}}_{{\bf{k}}}-\mu_0 & -s\lambda_{\text{SOC}} \\
-i\lambda_{\text{SOC}} & -s\lambda_{\text{SOC}} & \epsilon^{\text{xy}}_{{\bf{k}}}-\mu_0 \\
\end{pmatrix},
\end{equation}
with $s=\pm$ for the spin and
\begin{equation}
\begin{aligned}
& \epsilon^{\text{xy}}_{{\bf{k}}} = -2t_1\cos(k_x)-2t_2\cos(k_y),	\\
& \epsilon^{\text{yz}}_{{\bf{k}}} = -2t_2\cos(k_x)-2t_1\cos(k_y),	\\
& \epsilon^{\text{xz}}_{{\bf{k}}} = -2t_3(\cos(k_x)+\cos(k_y)) -4t_4\cos(k_x)\cos(k_y) \\
&\qquad  - 2t_5(\cos(2k_x)+\cos(2k_y)),	\\
& \epsilon^{\text{off}}_{{\bf{k}}} = -4t_6\sin(k_x)\sin(k_y).
\end{aligned}
\end{equation}
The $H_z$ term describes the hopping along $z$-direction and is introduced to deal with out-of-plane pairing such as ($d_{xz},d_{yz}$). Under the same basis $\psi_{s}({\bf{k}})$, it takes the form, 
\begin{equation}
H_z({\bf{k}}) = -8t_0 \cos(k_x/2)\cos(k_y/2)\cos(k_z/2).
\end{equation}
The best ARPES fit yields $[t_1,t_2,t_3,t_4,t_5,t_6,\mu_0,\lambda_{\text{SOC}}]=[0.145,0.016,0.081,0.039,0.05,0,0.122,0.032]$ eV \cite{Zabolotnyy2013}. We choose $t_0=0.01$ eV so that $t_0/t_1$ agrees with previous study \cite{Pavarini2006}. It should be noted that we have only considered $\bf{k}$-independent SOC, whose magnitude is consistent with density functional theory (DFT) calculations \cite{Clepkens2021}. It has been shown previously that including different types of ${\bf k}$-dependent SOC may enhance certain particular pairing states, but to make them dominant requires the ${\bf k}$-dependent SOC to be at least one order higher in magnitude than those predicted by DFT \cite{Suh2020,Clepkens2021}. Therefore, we will not discuss such possibility in this work. The resulting 3D Fermi surfaces of our model are plotted in Fig. \ref{fig1}(c), and the 2D orbital-resolved band structures are shown in Fig. \ref{fig1}(d) along a high symmetry line within the $k_z=0$ plane of the Brillouin zone. 

The above TB Hamiltonian also allows us to get a feeling about leading multipole fluctuations, which cannot be obtained currently from experiment \cite{Sidis1999,Servant2000,Steffens2019}. As shown in Appendix B, calculations based on random phase approximation (RPA) for $j=3/2$ yield two leading multipole correlations, $\langle \hat J_z\hat J_z\rangle$, $\langle \hat T_{ra}\hat T_{ra}\rangle$, in the AFM channel and four leading correlations, $\langle\hat J_z\hat J_z\rangle$, $\langle\hat T_{ra}\hat T_{ra}\rangle$, $\langle\hat T_{ra}\hat T_{rb}\rangle$ and $\langle\hat T_{rb}\hat T_{rb}\rangle$, in the FM channel. By contrast, electric multipole fluctuations are nearly unchanged for the  Stoner factor $\alpha_S<1$, implying that there is no electric instability. The definition of the Stoner factor is given in Appendix B. The leading multipoles are supposed to dominate the pairing interaction in the RPA approximation, but other components also should be present and have substantial contributions. Quite often, RPA cannot give a proper description of multipole fluctuations in real materials with strong electronic correlations. To have a systematic analysis of the electrons' pairing, we disentangle all multipole components allowed by symmetry and assume an empirical form of the interaction vertex \cite{Millis1990,Monthoux1991}: 
\begin{equation}
V^{j_1j_2\Gamma}({\bf{q}},i\nu_n) = \frac{1}{1+[{\bm{\xi}}\cdot({\bf{q}}-{\bf{Q}})]^2+|\nu_n|/\omega_{\bf q}},
\label{vertex}
\end{equation}
where $\nu_n$ is the bosonic Matsubara frequency, $\bm{\xi}=(\xi_{xy},\xi_{xy},\xi_{z})$ is the anisotropic correlation length of corresponding multipole fluctuations, $\omega_{\bf q}$ is the fluctuation energy, and ${\bf{Q}}$ is the characteristic wave vector for AFM, FM, or electric fluctuations. These parameters may in principle vary with $j_1$, $j_2$ and $\Gamma$. Here we drop the labels for simplicity. 

The above empirical form has the advantage to directly incorporate some important  information of the AFM, FM, or electric fluctuations from experiments. It reflects the  dynamical fluctuations of pairing interactions beyond mean-field approximation, which is often important in strongly correlated superconductors. Its exact form was initially proposed for the spin-fluctuation spectrum in cuprates and may be derived from a straightforward expansion of the momentum-dependent spin interaction \cite{Millis1990,Monthoux1991,Monthoux1992}. It was later applied to other unconventional superconductors and has explained many of their important properties \cite{Varma2012,Yang2014,Li2017,Li2018,Li2019,Liu2019,YLi2021}. Here we further extend it to multipole fluctuations in Sr$_2$RuO$_4$ in the presence of SOC. This inevitably introduces many free parameters, so we will study each multipole fluctuation individually before proposing a form of their mixture to simplify the discussion on potential pairing states in reality.

\section{Eliashberg equations}
Candidate pairing symmetries of the superconductivity can be analyzed using the linearized Eliashberg equations \cite{Monthoux1992,Li2018}:
\begin{equation}
\begin{aligned}
Z_{\mu}({\bf{k}},i\omega_n) = & 1 + \frac{\pi T}{\omega_n} \sum_{\mu',n'} \oint_{\text{FS}_{\mu'}}\frac{\text{d}{\bf{k}}'}{(2\pi)^3 v_{{\bf{k}}'_{\text{F}}}}\text{sgn}(\omega_{n'})
\\
& \times K^{\text{N}}_{\mu\mu'}({\bf{k}},i\omega_{n};{\bf{k}}',i\omega_{n'}),\\
\lambda \psi_{\mu}({\bf{k}},i\omega_n) = & \pi T\sum_{\mu',n'} \oint_{\text{FS}_{\mu'}}\frac{\text{d}{\bf{k}}'}{(2\pi)^3 v_{{\bf{k}}'_{\text{F}}}}\psi_{\mu'}({\bf{k}}',i\omega_{n'})
\\
& \times \frac{K^{\text{A}}_{\mu\mu'}({\bf{k}},i\omega_{n};{\bf{k}}',i\omega_{n'})}{|\omega_{n'}Z_{\mu'}({\bf{k}}',i\omega_{n'})|},
\label{gap}
\end{aligned}
\end{equation}
where $\omega_{n}$ and $\omega_{n'}$ denote the fermionic Matsubara frequencies, $\mu$ and $\mu'$ are band indices, the integral with FS$_{\mu'}$ is over the Fermi surface of band $\mu'$ with corresponding Fermi velocity $v_{{\bf{k}}'_{\text{F}}}$, $Z_{\mu}$ is the renormalization function, and $\psi_{\mu}=\Delta_{\mu}Z_{\mu}$ is the anomalous self-energy related to the gap function $\Delta_\mu$. All bands are doubly degenerate with pseudospin $\eta=\pm$ and we only consider intraband pairing (singlet or triplet over pseudospin). Figure \ref{fig1}(b) shows the Feynman diagram for the anomalous self-energy $\psi_\mu$. During our calculations, the kernel functions $K^{\text{N}}_{\mu\mu'}$ and $K^{\text{A}}_{\mu\mu'}$ can be determined from the interacting Hamiltonian $H_{\text{int}}$ using the above empirical pairing interactions and take the following forms:
\begin{widetext}
\begin{equation}
\begin{aligned}
K^{\text{N}}_{\mu\mu'}({\bf{k}},i\omega_{n};{\bf{k}}',i\omega_{n'}) = &\sum_{\substack{j_1j_2\Gamma \\ \alpha\beta}} \sum_{\substack{lml'm' \\ \eta\eta'}} 
g^{j_1j_2\Gamma}_{\alpha\beta} V^{j_1j_2\Gamma}({\bf{k}}-{\bf{k}}',i\omega_{n}-i\omega_{n'})\mathrm{Re}[Q_{ml}^{j_1\Gamma\alpha*}Q_{l'm'}^{j_2\Gamma\beta} u_{j_1l,\mu\eta}^{{\bf{k}}*}u_{j_1m,\mu'\eta'}^{{\bf{k}}'}u_{j_2l',\mu'\eta'}^{{\bf{k}}'*}u_{j_2m',\mu\eta}^{{\bf{k}}}],\\
K^{\text{A}}_{\mu\mu'}({\bf{k}},i\omega_{n};{\bf{k}}',i\omega_{n'}) = &
\sum_{\substack{j_1j_2\Gamma \\ \alpha\beta}} \sum_{\substack{lml'm' \\ \eta}} 
g^{j_1j_2\Gamma}_{\alpha\beta}Q_{ml}^{j_1\Gamma\alpha*}Q_{l'm'}^{j_2\Gamma\beta} [ V^{j_1j_2\Gamma}({\bf{k}}-{\bf{k}}',i\omega_{n}-i\omega_{n'}) u_{j_1l,\mu\eta}^{{\bf{k}}*} u_{j_1m,\mu'\eta}^{{\bf{k}}'} u_{j_2l',\mu\bar\eta}^{-{\bf{k}}*} u_{j_2m',\mu'\bar\eta}^{-{\bf{k}}'}
\\
& + V^{j_1j_2\Gamma}({\bf{k}}+{\bf{k}}',i\omega_{n}+i\omega_{n'}) u_{j_1l,\mu\eta}^{{\bf{k}}*} u_{j_1m,\mu'\bar\eta}^{-{\bf{k}}'} u_{j_2l',\mu\bar\eta}^{-{\bf{k}}*} u_{j_2m',\mu'\eta}^{{\bf{k}}'}],
\end{aligned}
\end{equation}
\end{widetext}
where $\hat u^{{\bf{k}}}$ is the matrix diagonalizing the 3D or 2D TB Hamiltonian, projected in the $j$ representation. The linearized Eliashberg equations can be solved numerically by approximating $\Delta_{\mu}({\bf{k}})\equiv\Delta_{\mu}({\bf{k}},i\omega_{n})\approx\Delta_{\mu}({\bf{k}},i\pi T_c)$ and using 1024 Matsubara frequencies, $41\times41\times41$ ${\bf{k}}$ meshes in the 3D Brillouin zone or $201\times201$ ${\bf{k}}$ meshes in the 2D Brillouin zone. Each eigenvector $\lambda$ of Eq. (\ref{gap}) corresponds to a single solution of electron pairing and gives the corresponding gap structure on the Fermi surfaces. The largest eigenvalue $\lambda$ of $\psi_\mu$ at $T_c$ determines the leading pairing state.

\section{Individual pairing interactions}
The results of individual multipole fluctuation channels are presented in Fig.~\ref{fig2} for $j=3/2$ and Figs. \ref{fig7} and \ref{fig8} in Appendix D for $j=5/2$ and $j$-mixed subspaces, respectively. We compare the eigenvalues of six major pairing states, $s$, $d_{x^2-y^2}$, $(p_x,p_y)$, $g$, $d_{xy}$, $(d_{xz},d_{yz})$, for the 3D Hamiltonian $H_{\text{3D}}$ of Sr$_2$RuO$_4$. The parameter $g^{j_1j_2\Gamma}_{\alpha\beta}$ is chosen such that all multipole fluctuations are treated equally for each set of calculations.

\begin{figure}[t]
\centering
\includegraphics[width=8.6 cm]{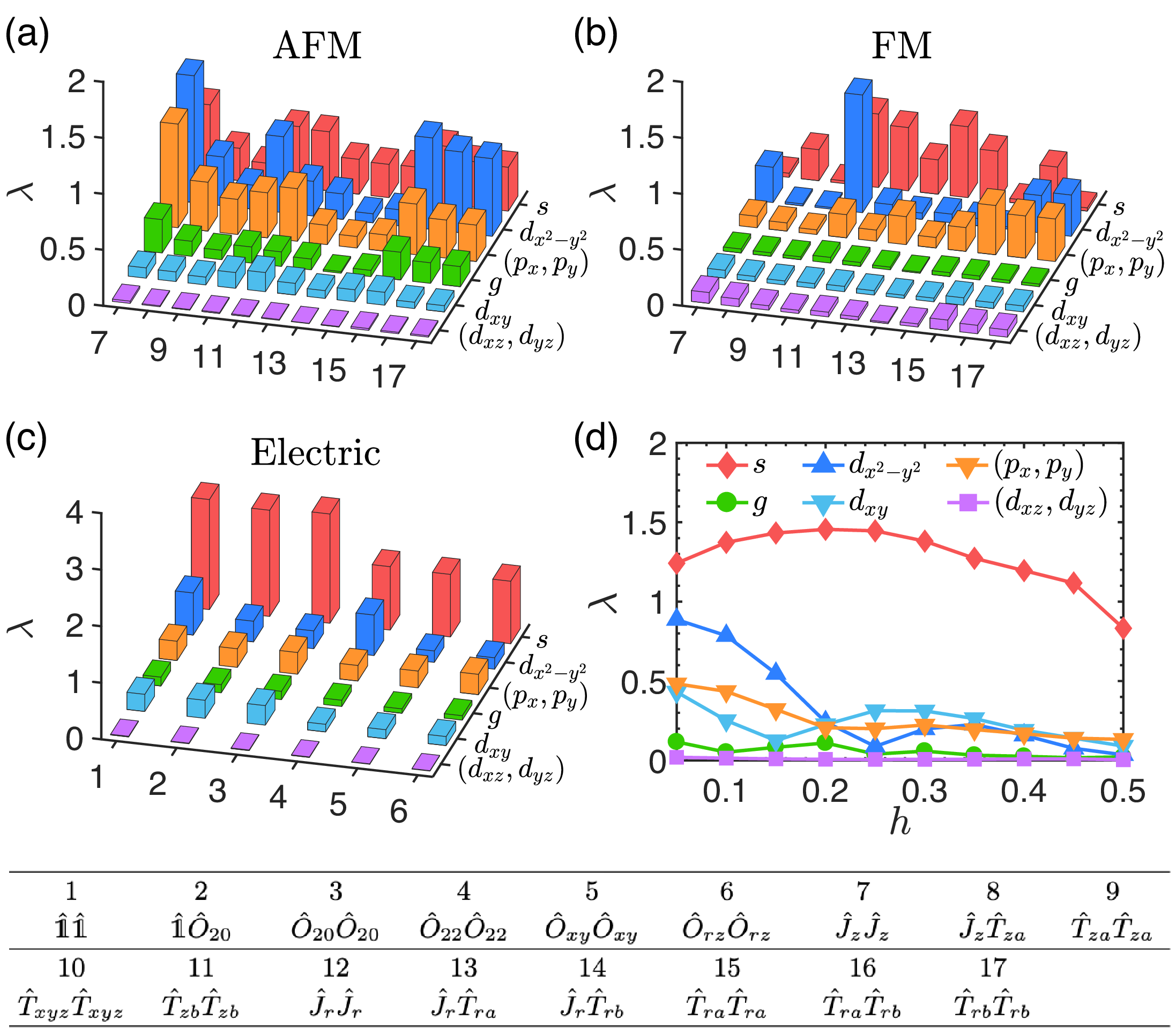}
\caption{Eigenvalues of six major pairing states, $s$ ($A_{1g}$), $d_{x^2-y^2}$ ($B_{1g}$), $(p_x,p_y)$ ($E_u$), $g$ ($A_{2g}$), $d_{xy}$ ($B_{2g}$), and $(d_{xz},d_{yz})$ ($E_g$), for individual pairing interactions in $j=3/2$ from (a) 11 AFM multipole channels with $\xi^{\text{AFM}}_{xy}=9.7$ {\AA}, $\omega_{\bf q}=\omega^{\text{AFM}}_0=11.1$ meV, and ${\bf{Q}}_{\text{AFM}}=(0.3,0.3,0)$; (b) 11 FM multipole  channels with $\xi^{\text{FM}}_{xy}=2.5$ {\AA}, $\omega_{\bf q}=v_0|{\bf q}|$, and ${\bf{Q}}_{\text{FM}}=(0,0,0)$; (c) 6 electric multipole fluctuation channels with $\xi^{\text{E}}_{xy}=\xi^{\text{AFM}}_{xy}$, $\omega_{\bf q}=\omega_0^{\text{E}}=\omega_0^{\text{AFM}}$, and ${\bf{Q}}_{\text{E}}=(0.2,0.2,0)$. (d) Eigenvalues of six major pairing states for averaged electric multipole fluctuations as a function of ${\bf{Q}}_{\text{E}}$ along $(h,h,0)$ direction. The $s$-wave state always dominates and has a maximal eigenvalue around ${\bf{Q}}_{\text{E}}=(0.2,0.2,0)$. The table on the bottom lists all multipole fluctuation components for $j=3/2$, sorted according to their IRs and ranks. The longitudinal correlation length is set to $\xi_{z}=0.1\xi_{xy}$.}
\label{fig2}
\end{figure}

For AFM fluctuations, inelastic neutron scattering (INS) experiments estimate $\xi^{\text{AFM}}_{xy}=9.7$ {\AA} and $\omega_{\bf q}=\omega^{\text{AFM}}_0=11.1$ meV at the AFM wave vector ${\bf{Q}}_{\text{AFM}}=(0.3,0.3,q_l)$ \cite{Sidis1999,Servant2000,Steffens2019}, where $q_l$ will be set to zero in our calculations but its value shows no qualitative influence on our results. The longitudinal correlation length is set to $\xi^{\text{AFM}}_z=0.1\xi^{\text{AFM}}_{xy}$ to reflect the absence of $z$-axis signal \cite{Servant2000}. As shwon in Figs.~\ref{fig2}(a), among all 11 AFM multipole fluctuation channels for $j=3/2$, $d_{x^2-y^2}$ or $s$ are most supported. Two leading fluctuation channels from RPA analysis, $\hat J_{z}\hat J_{z}$ and $\hat T_{ra}\hat T_{ra}$, gives predominant $d_{x^2-y^2}$-wave pairing. The subordinate channels, $\hat J_{z}\hat T_{za}$, $\hat T_{xyz}\hat T_{xyz}$, $\hat T_{ra}\hat T_{rb}$, $\hat T_{rb}\hat T_{rb}$, also support $d_{x^2-y^2}$, while the subordinate $\hat T_{zb}\hat T_{zb}$, $\hat J_r\hat J_r$, $\hat J_r\hat T_{ra}$, $\hat J_r\hat T_{rb}$ favor $s$-wave and $\hat T_{za}\hat T_{za}$ favors $(p_x,p_y)$ or $p_x+ip_y$. For $j=5/2$ as shown in Fig.~\ref{fig7}(a), the leading dipole component $\hat J_{z}\hat J_{z}$ supports $d_{x^2-y^2}$, while the leading dotriacontapole $\hat D_{za2}\hat D_{za2}$ supports $s$-wave pairing. Figure~\ref{fig8}(a) gives the results for $j$-mixed AFM fluctuations. We see $s$ and $d_{x^2-y^2}$ are also two most favored pairing states than others.

FM pairing interactions have previously been considered because Sr$_2$RuO$_4$ has similar electronic structures as the itinerant ferromagnets SrRuO$_{3}$ and Sr$_{4}$Ru$_{3}$O$_{10}$ and the metamagnet Sr$_{3}$Ru$_{2}$O$_{7}$ \cite{Bergemann2003,Mackenzie2017}. A short-range FM order was reported in experiment in Co-doped Sr$_2$RuO$_4$, indicating that Sr$_2$RuO$_4$ might be near a FM instability \cite{Ortmann2013}. PNS experiment in Sr$_2$RuO$_4$ has also reported a broad FM response \cite{Steffens2019}, giving ${\bf{Q}}_{\text{FM}}=(0,0,0)$, $\xi^{\text{FM}}_{xy}=2.5$ {\AA}, and a characteristic energy $\omega_0^{\text{FM}}=15.5$ meV. Since there is no further experimental details for the FM multipole fluctuations, we will take simply $\omega_{\bf q}=v_0|{\bf q}|$ and choose $v_0$ such that $\omega_{\bf q}$ reaches the order of $\omega_0^{\text{FM}}$ at the zone boundary. A slight variation of $v_0$ makes no qualitative change on our main conclusions. Figure \ref{fig2}(b) shows the typical results of six major pairing states induced by FM pairing interactions for $j=3/2$. Similarly, we find predominant $d_{x^2-y^2}$-wave pairing from leading dipole fluctuations $\hat J_{z}\hat J_{z}$ and ($p_x,p_y$)-wave from leading octupole $\hat T_{ra}\hat T_{ra}$, $\hat T_{ra}\hat T_{rb}$, and $\hat T_{rb}\hat T_{rb}$, while the $s$-wave is supported by some subordinate multipole channels. For $j=5/2$ in Fig.~\ref{fig7}(b), the leading dipole $\hat J_{z}\hat J_{z}$ favors $s$-wave, while the leading dotriacontapole $\hat D_{ra1}\hat D_{ra1}$ and $\hat D_{rb}\hat D_{rb}$ support ($p_x,p_y$)-wave. In Fig.~\ref{fig8}(b), $s$ and $d_{x^2-y^2}$ are supported by most $j$-mixed FM channels.

Electric fluctuations may arise from multi-orbital nature of Sr$_2$RuO$_4$ \cite{Raghu2010,Mravlje2011,Acharya2017,Boehnke2018,Acharya2019} and have similar interaction vertex as AFM ones, but with ${\bf{Q}}_{\text{E}}=(0.2,0.2,0)$, $\xi^{\text{E}}_{xy}=\xi^{\text{AFM}}_{xy}$, and $\omega_0^{\text{E}}=\omega_0^{\text{AFM}}$. As shown in Fig. \ref{fig2}(c), all six multipole channels for $j=3/2$ support $s$-wave pairing, which is robust under the tuning of $\xi^{\text{E}}_{xy}$ and $\omega_0^{\text{E}}$. Figure \ref{fig2}(d) plots the eigenvalues of six major pairing states as a function of ${\bf{Q}}_{\text{E}}$ along the (110) direction. We see that the $s$-wave pairing always has a much larger eigenvalue than others. This is expected since superconductivity induced by charge fluctuation is typically $s$-wave. But quite interestingly, the eigenvalue of $s$ reaches a maximum around ${\bf{Q}}_{\text{E}}=(0.2,0.2,0)$, exactly the wave vector proposed by the RPA charge susceptibility \cite{Acharya2019}, implying a potential role of electric multipole fluctuations in superconducting Sr$_2$RuO$_4$. For $j=5/2$ and $j$-mixed subspaces, Figs.~\ref{fig7}(c) and \ref{fig8}(c) show that the $s$-wave pairing is always supported by leading electric multipole fluctuations.

All together, $s$ and $d_{x^2-y^2}$ are most supported by individual multipole pairing interactions, while the rest four are less favored. Thus, to discuss the possibility of $g$ and other three pairing states, it is necessary to go beyond individual multipole channels and consider some mixed form, which is reasonable given their coexistence. To proceed, we first note that all but $(d_{xz},d_{yz})$ are $m_z$-symmetric, i.e., symmetric about the $k_z=0$ plane, whose relative importance can be evaluated in 2D. By contrast, the $m_z$-antisymmetric $(d_{xz},d_{yz})$ or $d_{xz}+id_{yz}$ pairing state requires 3D calculations as performed above, but it is not favored because our TB Hamiltonian $H_{\text{3D}}$ is only weakly dispersive along $k_z$ direction. We find this conclusion quite robust against reasonable tuning of $t_0$ and $\xi_z$ within our framework. As an example, we give the results with an artificially enhanced $\xi_z=\xi_{xy}$ for $j=3/2$ in Appendix C. In the literature, it has been proposed that $d_{xz}+id_{yz}$ may become dominant if a sizable ${\bf k}$-dependent SOC of $E_g$ representation is included \cite{Suh2020}, which induces interorbital hopping along $z$-direction but has to be more than one order of magnitude higher than that of DFT prediction \cite{Clepkens2021}. Hence, we will no longer consider $d_{xz}+id_{yz}$ in the following section to reduce computational efforts by performing calculations only in 2D.

\section{Mixed pairing interactions}
From now on, we will focus on the 2D Hamiltonian $H_{\text{2D}}$. In real materials, different multipole fluctuations may coexist, so we must consider their possible combinations, namely, a mixed pairing interaction of AFM, FM, and electric multipole fluctuations such as 
\begin{equation}
V_{\text{mix}} = r_1V^{\text{AFM}} + r_2 V^{\text{FM}} + r_3 V^{\text{E}},
\end{equation}
where $V^{\text{AFM}}$, $V^{\text{FM}}$, and $V^{\text{E}}$ denote the AFM, FM, and electric multipole pairing interactions, respectively, and $r_i$ controls their relative strength. There may exist different combinations of individual multipole channels, but a natural one without much priori knowledge is to average each term over all multipole components in their respective $j=3/2$, $5/2$, or $j$-mixed subspaces. This work will focus on this particular form of the pairing interaction. But to examine the robustness of our results for other possible combinations, we also show in Appendix E the results for pairing interactions averaged only over $j=3/2$ or over both $j=3/2$ and 5/2 subspaces. All other parameters are set to be the same as discussed in previous section.

\begin{figure}
\begin{center}
\includegraphics[width=8 cm]{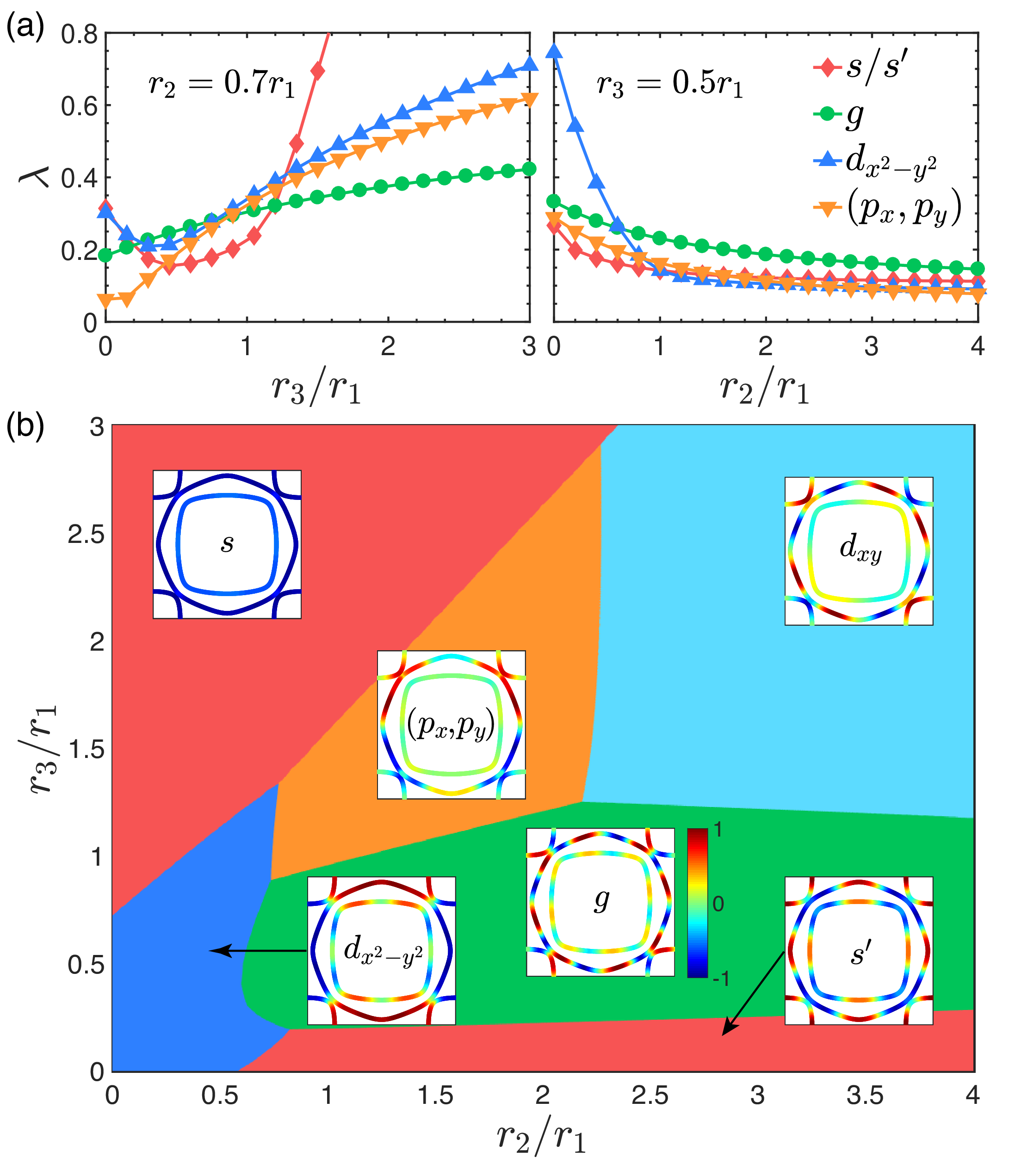}
\caption{(a) Eigenvalues of $s/s'$, $g$, $d_{x^2-y^2}$ and $(p_x,p_y)$ pairing states as a function of the ratio $r_3/r_1$ at $r_2/r_1=0.7$ and $r_2/r_1$ at $r_3/r_1=0.5$ for a mixed AFM, FM, and electric pairing interaction. (b) Theoretical phase diagram of predominant pairing states on the plane of $r_2/r_1$ and $r_3/r_1$. The insets show corresponding gap structures in  each region.}
\label{fig3}
\end{center}
\end{figure}

The resulting phase diagram including $j=3/2$, 5/2, and $j$-mixed terms is plotted in Fig. \ref{fig3}, together with two examples of the largest eigenvalues of four major pairing states and their variation with the ratio $r_2/r_1$ or $r_3/r_1$. The $d_{x^2-y^2}$-wave is seen to extend from the origin  ($r_2=r_3=0$) to cover a major part of the phase diagram with dominant AFM multipole fluctuations ($r_2/r_1<0.6$, $r_3/r_1<0.7$). A nodal $s$-wave is induced by a moderate FM pairing interaction ($r_2/r_1>0.6$), but for a stronger electric interaction ($r_3/r_1>0.7$), we find a predominant nodeless $s$-wave. For distinction, we will use $s'$ to denote nodal $s$-wave in the following. As discussed earlier, $d_{x^2-y^2}$ and $s$ (or $s'$) are major pairing states for dominant AFM, FM or electric multipole fluctuations. In the absence of electric fluctuations ($r_3=0$), the resulting $d_{x^2-y^2}$ or $s'$ waves agree well with previous work \cite{Romer2019}. Quite surprisingly, $g$-wave pairing is seen in Fig. \ref{fig3} to govern a large portion of the phase diagram where both FM and electric multipole fluctuations are of equal importance as the AFM ones.

Thus, within our framework, the accidentally degenerate $d_{x^2-y^2}+ig$ pairing may exist at the phase boundary with a somewhat weaker FM pairing interaction than the AFM one, namely $r_2/r_1\approx 0.6$. However, a moderate electric pairing interaction ($0.2<r_3/r_1<0.8$) is also needed. If the electric fluctuations are too weak or too strong, a two-component $s'+id_{x^2-y^2}$ or $s+id_{x^2-y^2}$ might appear but is inconsistent with the ultrasound experiments \cite{Benhabib2021,Ghosh2021}. The other two states, $(p_{x},p_{y})$ and $d_{xy}$, require even stronger FM or electric fluctuations than AFM ones. In all cases, electric multipole fluctuations, such as 0-rank charge fluctuations, seem to play a crucial role in superconducting Sr$_2$RuO$_4$, and should be better examined by future X-ray diffraction or Raman experiments \cite{Feng2012,Croft2014,Gallais2013,Xi2015,Thorsmolle2016}.

The emergence of $g$-wave is robust for such mixed AFM, FM, and electric pairing interactions. Appendix E shows the phase diagrams of two other examples of averaged pairing interactions. In the first case, we consider only the $j=3/2$ subspace and take an averaged pairing interaction over all multipole fluctuations; and in the second case, we take the average over both $j=3/2$ and 5/2 subspaces. The phase diagrams are shown in Fig.~\ref{fig9} and found qualitatively the same. The downshift of the phase boundaries in Fig. \ref{fig3}(b) indicates that the $s$-wave pairing is more promoted by the $j$-mixed multipole interaction. Nevertheless, the $g$-wave state still covers a large portion of the phase diagram with strong AFM fluctuations and moderate FM and electric ones, where $d_{x^2-y^2}$ and $s$ (or $s'$) solutions supported by individual multipole fluctuations are suppressed. We thus conclude that the competition and interplay of AFM, FM, and electric multipole fluctuations may provide a mechanism for the unusual $d_{x^2-y^2}+ig$ pairing in Sr$_2$RuO$_4$. Whether or not this reflects the true situation in real material requires future experimental scrutiny.

\begin{figure}[t]
\begin{center}
\includegraphics[width=8.2 cm]{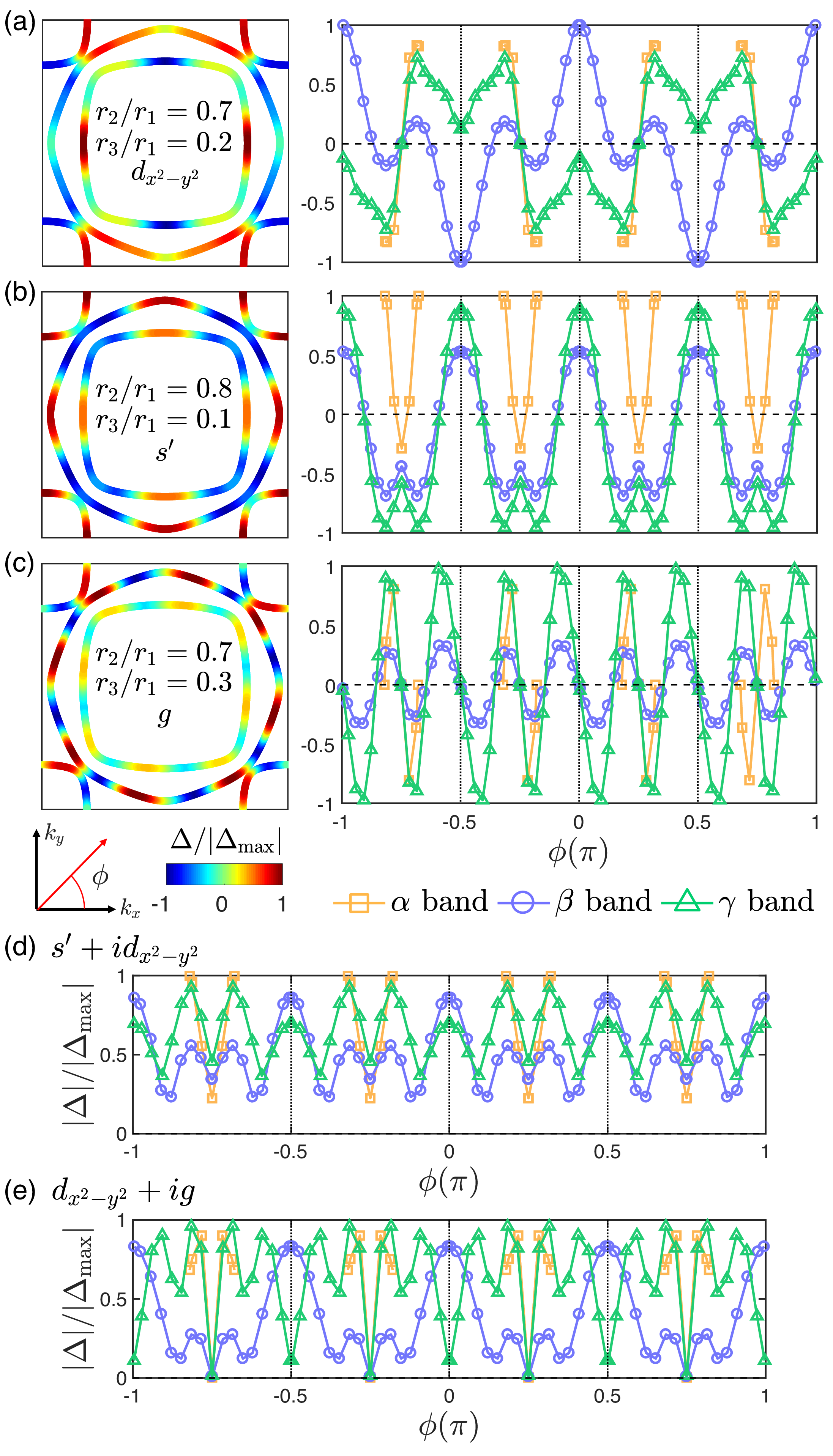}
\caption{Typical gap structures of (a) $d_{x^2-y^2}$, (b) $s'$, and (c) $g$-wave pairing states and their evolutions with the azimuth $\phi$. (d) Gap magnitude of a typical $s'+id_{x^2-y^2}$ pairing state constructed from (a,b) as a function of the azimuth $\phi$, showing local gap minima along the zone diagonal ($\phi=\pm \pi/4$ and $\pm 3\pi/4$) on all three bands. (e) Gap magnitude of a typical $d_{x^2-y^2}+ig$ pairing state from (b,c) as a function of the azimuth $\phi$, showing the symmetry-protected nodes along the zone diagonal.}
\label{fig4}
\end{center}
\end{figure}

To have an idea of the gap structures for these pairing states, Fig. \ref{fig4} presents their projection (normalized) on the 2D Fermi surfaces and evolution with the azimuth $\phi$. Both $d_{x^2-y^2}$ and $g$ show clear symmetry-protected nodes on three bands along the zone diagonal ($\phi=\pm \pi/4$ and $\pm 3\pi/4$). The resulting $d_{x^2-y^2}+ig$ gap has also nodes along the zone diagonal, which is protected by symmetry and fits well the STM data \cite{Sharma2020}. Quite unexpectedly, we also find that the $s'$-wave can change sign or have gap minima near the zone diagonal. This interesting feature arises from the particular orbital character of the three bands. Along the diagonal direction, $\alpha$ band contains contribution only from $|j=\frac32,j_z=\pm\frac32\rangle$ and $|\frac52,\pm\frac52\rangle$, $\beta$ band contains only $|\frac52,\pm\frac32\rangle$ and $|\frac52,\pm\frac52\rangle$, and the $\gamma$ band contains only $|\frac32,\pm\frac12\rangle$ and $|\frac52,\pm\frac12\rangle$. Hence, as an example, an $s$-wave pairing supported by multipole fluctuations of $j=3/2$, $j_z=\pm 1/2$, or $j_z=\pm3/2$ must have nodes along the zone diagonal on $\beta$, $\alpha$, or $\gamma$ band, respectively. This may give rise to the gap minima after other contributions are included. Consequently, the two-component $s'+id_{x^2-y^2}$ can also exhibit gap minima near the zone diagonal. If we assume the two components have the same magnitude, we may obtain a gap structure in Fig.~\ref{fig4}(d), where the relative gap ratio is $|\Delta_{\pi/4}/\Delta_{\text{max}}|\approx0.22$, 0.45, 0.34 for $\alpha$, $\beta$, $\gamma$ bands, respectively. Note that previous STM experiment has an energy resolution of about 75 $\mu$eV, which is roughly 21\% of the measured gap of 350 $\mu$eV \cite{Sharma2020}. Thus, it is impossible for the STM alone to exclude $s'+id_{x^2-y^2}$ if the $s'$ component is only moderate compared to $d_{x^2-y^2}$ component. The ultrasound experiment is then crucial.

\section{Discussion and Conclusions}
Our results provide a potential physical basis for the possibility of $d_{x^2-y^2}+ig$ pairing in superconducting Sr$_2$RuO$_4$. But there are also proposals supporting other pairing states in the literature. For clarity, we give a brief summary in this section on current experimental and theoretical situation of several major candidates of two-component order parameter that breaks the time reversal symmetry, including $p_x+ip_y$, $d_{xz}+id_{yz}$, $s'+id_{x^2-y^2}$, $s+id_{xy}$, and $d_{x^2-y^2}+ig$.

The purely odd-parity $p_x+ip_y$ paring was recently excluded by a series of NMR \cite{Pustogow2019,Ishida2020,Chronister2021} and PNS \cite{Petsch2020} experiments. A mixed-parity state has been proposed in the quasi-1D limit \cite{Scaffidi2020}, which has accidental nodes along the zone diagonal consistent with the STM experiment \cite{Sharma2020}. However, the odd-parity component requires a jump in the shear elastic modulus $(c_{11}-c_{12})/2$, which was not observed in the ultrasound experiment, at least within the current accuracy \cite{Ghosh2021,Benhabib2021}. In our theory, the $p_x+ip_y$ paring is only supported by several ferromagnetic fluctuation components such as the octupole $\hat T_{ra}\hat T_{ra}$, $\hat T_{ra}\hat T_{rb}$, $\hat T_{rb}\hat T_{rb}$ of $j=3/2$, and therefore only appears for relatively strong FM and electric fluctuations in the unphysical region of $r_2/r_1\approx1-2$ and $r_3/r_1\approx1-3$.

The $d_{xz}+id_{yz}$ pairing is typically unfavored due to the quasi-2D Fermi surface topology of Sr$_2$RuO$_4$, but may be stabilized if a sizable momentum-dependent $E_g$-SOC is included \cite{Suh2020}. The latter could give rise to a spin-triplet odd-orbital $(d_{xz}+id_{yz})$-like pairing that can explain the Knight shift drop below $T_c$ \cite{Yu2018,Lindquist2020,Fukaya2022}, but the required $E_g$-SOC is at least one order of magnitude higher than that predicted by DFT \cite{Clepkens2021}. The $d_{xz}+id_{yz}$ gap is featured with horizontal line nodes on the $k_z=0$ plane \cite{Ramires2021}, so it is supposed to cause spin resonance at $q_l=0$, which is, however, absent according to recent neutron scattering experiments \cite{Iida2020,Jenni2021}. It is also inconsistent with the ultrasound experiment showing no evident jump in $(c_{11}-c_{12})/2$ \cite{Ghosh2021,Benhabib2021}. The $d_{xz}+id_{yz}$ pairing was mainly supported by $\mu$SR measurements that reported the splitting of superconductivity and TRSB under uniaxial pressure as opposed to hydrostatic pressure \cite{Grinenko2021a,Grinenko2021b}. However, this splitting was questioned by specific heat measurements, which found no sign of bulk phase transition induced by uniaxial pressure \cite{Li2021}. More accurate experiments will be needed to clarify how exactly superconductivity evolves under pressure. In our calculations, the SOC is ${\bf k}$-independent and has a magnitude consistent with DFT prediction. Thus, the $d_{xz}+id_{yz}$ pairing is always unfavored within our framework.

The $d_{x^2-y^2}$-wave has the desired vertical line nodes revealed by thermal conductivity \cite{Hassinger2017} and nodes or gap minima on $\alpha$ and $\beta$ bands in STM measurements \cite{Sharma2020}. From our calculations, it is indeed supported by AFM fluctuations and can form a two-component order parameter with accidentally degenerate $s'$ or $g$ in the presence of moderate FM and electric fluctuations. An $s'+id_{x^2-y^2}$ has been proposed in previous theory but was nodeless along the zone diagonal \cite{Romer2019}. By contrast, our derived $s'+id_{x^2-y^2}$ can have nodes or gap minima near the 2D zone diagonal and may agree with STM. But $s'+id_{x^2-y^2}$ seems inconsistent with ultrasound experiment, where the observed thermodynamic jump of the shear elastic modulus $\delta c_{66}\propto\alpha_4^2$ reflects the coupling term $\alpha_4 u_{xy}(\Delta_{s'}^*\Delta_{d_{x^2-y^2}}+\Delta_{d_{x^2-y^2}}^*\Delta_{s'})$ between the strain $u_{xy}$ and two superconducting components in the Landau free energy \cite{Benhabib2021,Ghosh2021}. Such a coupling is prohibited by symmetry because $B_{2g}(u_{xy})\otimes A_{1g}(\Delta_{s'})\otimes B_{1g}(\Delta_{d_{x^2-y^2}})=A_{2g}\not=A_{1g}$. 

To overcome this problem, an accidentally degenerate $s'+id_{xy}$ pairing has been proposed by taking into consideration the nearest-neighbor Coulomb repulsion \cite{Romer2021}, which is nodeless on $\alpha$ band but has gap minima on $\gamma$ band along the azimuthal $\phi=0.15\pi$ direction. A recent analysis suggested that this $s'+id_{xy}$ gap structure could well fit the Bogoliubov quasiparticle interference pattern in STM measurement \cite{Bhattacharyya2021} and might be a competitive candidate for the pairing symmetry of superconducting Sr$_2$RuO$_4$. Calculations of the spin and charge susceptibilities indicated that the primary role of nearest-neighbor Coulomb repulsion is to enhance the electric fluctuations over the magnetic ones \cite{Romer2021}. It thus corresponds to increase $r_3/r_1$ in our theory. In this sense, their result is consistent with our phase diagram, where $d_{xy}$ can indeed become dominant at large $r_3/r_1>1$. But within our framework, it also requires very strong FM fluctuations ($r_2/r_1>2$), which is not realistic in experiment \cite{Steffens2019}.

An interorbital spin-triplet $d_{x^2-y^2}+ig$ state has been proposed by including a sizable momentum-dependent $B_{2g}$-SOC about 20 times higher than that predicted by DFT \cite{Clepkens2021}. It is different from our pseudospin singlet $d_{x^2-y^2}+ig$ solution which is a mixture of spin-singlet even-orbital and spin-triplet odd-orbital pairings but dominated by the spin-singlet component with a ${\bf k}$-independent SOC of reasonable magnitude. In any case, $d_{x^2-y^2}+ig$ has the desired nodal structures for STM and the required symmetry ($B_{2g}(u_{xy})\otimes B_{1g}(\Delta_{d_{x^2-y^2}})\otimes A_{2g}(\Delta_{g})=A_{1g}$) by ultrasound experiment, and may also find signatures in impurity scattering \cite{Hashitani2020,Zinkl2021} or heat capacity \cite{Wagner2021}. However, as for all accidentally degenerate pairing states, $d_{x^2-y^2}+ig$ also suffers from the difficulty in explaining the observed unsplitting of the transition under hydrostatic pressure \cite{Grinenko2021a,Grinenko2021b}, as well as the lack of a large specific heat jump at the TRSB transition under uniaxial pressure \cite{Li2021}. It has been argued that the unsplit transition under hydrostatic pressure can be accounted for by proper assumption of the Landau-Ginzburg parameters \cite{Yuan2021}, and a modified $d_{x^2-y^2}+ig$ scenario based on strain inhomogeneity near dislocations or domain walls may explain the lack of specific heat jump and the accidental degeneracy \cite{Yuan2021,Willa2021}. Unfortunately, these effects cannot be easily parameterized in our calculations, so that a quantitative justification of their analyses is not immediately possible. On the other hand, on a very crude level, one may expect that hydrostatic pressure tends to modify all parameters in the Hamiltonian, which may increase the bandwidth but keep the ratio $r_2/r_1$ and $r_3/r_1$ less affected. If this is the case, one may expect that the system could be driven away from magnetic instabilities, so that $T_c$ decreases but the superconductivity stays near the boundary of $d_{x^2-y^2}$ and $g$-waves with an unsplit transition. Under uniaxial strain, the system is driven towards an incommensurate spin-density-wave (SDW) instability as proposed in experiment \cite{Grinenko2021a}, which may primarily enhance (for small strain) AFM fluctuations or $r_1$. As shown in Fig. \ref{fig3}(a), starting from the $g$-dominant region, the eigenvalue of $d_{x^2-y^2}$ increases more rapidly with decreasing $r_2/r_1$ and $r_3/r_1$, while that of the $g$-wave varies only slightly. This would lead to an increase of $T_c$ in the $d_{x^2-y^2}$ pairing channel before it reaches a maximum near the SDW transition. At the same time, the $g$-wave component should remain less affected, which explains the splitting of the superconducting transition and the nearly unchanged $T_{\text{TRSB}}$ induced by the $g$-wave channel below $T_c$. In this sense, our calculations are consistent with the pressure experiments, but more elaborated analyses with realistic parameterization are needed for a final justification.

Putting together, $d_{x^2-y^2}+ig$ seems to be a most probable candidate for superconducting pairing in Sr$_2$RuO$_4$, at least within our framework. It may arise naturally from a mixed pairing interaction of AFM, FM, and electric multipole fluctuations of reasonable magnitudes. Our results provide a plausible explanation of the pairing symmetry in superconducting Sr$_2$RuO$_4$, and pose a challenge for future experiments to examine the role of different multipole fluctuations. Our theory may also be applied to other unconventional superconductors.
 
\acknowledgments
This work was supported by the National Natural Science Foundation of China (NSFC Grant No. 11974397, No. 12174429), the National Key Research and Development Program of MOST of China (Grant No. 2017YFA0303103), the Strategic Priority Research Program of the Chinese Academy of Sciences (Grant No. XDB33010100), the China Postdoctoral Science Foundation (Grant No. 2020M670422), and the Youth Innovation Promotion Association of CAS.

\appendix
\begin{table*}
\caption{\label{tab2}Definition of multipole operators in Table \ref{tab1} based on the operator-equivalent technique, which are classified according to the irreducible representations $\Gamma$ of $D_{4h}$ point group. The $j=5/2$ manifold contains operators from rank 0 to rank 5 (monopole $\mathds{1}$; dipole $J$; quadrupole $O$; octupole $T$; hexadecapole $H$; dotriacontapole $D$), while multipole operators in $j=3/2$ are up to rank 3 (monopole $\mathds{1}$; dipole $J$; quadrupole $O$; octupole $T$). For simplicity, we have used the same symbols for both $j$-spaces. They have in principle different bases and representation matrices.}
\renewcommand\arraystretch{1.5}
\centering
\begin{tabular}{ccccl}
\hline
\hline 
&\ \ \ \ IR($\Gamma$)\ \ \ \  & \ \ \ \ $\alpha$\ \ \ \  & \ \ \ \ $\hat Q^{j\Gamma\alpha}$\ \ \ \  & Basis \\
\hline  
\multirow{12}{*}{Electric} & \multirow{4}{*}{$A_{1g}^+$} & 1 & $\hat{\mathds{1}}$ & $\hat J_{00}$	\\
		& & 2 & $\hat O_{20}$ & $\hat J_{20}$	\\
		& & 3 & $\hat H_0$ & $(\sqrt{7}\hat J_{40}+\sqrt{5}\hat J_{44c})/\sqrt{12}$	\\
		& & 4 & $\hat H_4$ & $(\sqrt{5}\hat J_{40}-\sqrt{7}\hat J_{44c})/\sqrt{12}$	\\
\cline{2-5}
& $A_{2g}^+$ & 1 & $\hat H_{za}$ & $\hat J_{44s}$	\\
\cline{2-5}
& \multirow{2}{*}{$B_{1g}^+$} & 1 & $\hat O_{22}$ & $\hat J_{22c}$	\\
		& & 2 & $\hat H_2$ & $-\hat J_{42c}$	\\
\cline{2-5}
& \multirow{2}{*}{$B_{2g}^+$} & 1 & $\hat O_{xy}$ & $\hat J_{22s}$	\\
		& & 2 & $\hat H_{zb}$ & $\hat J_{42s}$	\\
\cline{2-5}
& \multirow{3}{*}{$E_{g}^+$} & 1 & $\hat O_{xz},\hat O_{yz}$ & $\hat J_{21c},\hat J_{21s}$	\\
	  & & 2 & $\hat H_{xa},\hat H_{ya}$ & $(-\hat J_{43c}+\sqrt{7}\hat J_{41c})/\sqrt{8},(\hat J_{43s}+\sqrt{7}\hat J_{41s})/\sqrt{8}$	\\
	  & & 3 & $\hat H_{xb},\hat H_{yb}$ & $(-\sqrt{7}\hat J_{43c}-\hat J_{41c})/\sqrt{8},(\sqrt{7}\hat J_{43s}-\hat J_{41s})/\sqrt{8}$	\\
\hline
\multirow{15}{*}{Magnetic} & $A_{1g}^-$ & 1 & $\hat D_4$ & $\hat J_{54s}$	\\
\cline{2-5}
& \multirow{4}{*}{$A_{2g}^-$} & 1 & $\hat J_z$ & $\hat J_{10}$	\\
& & 2 & $T_{za}$ & $\hat J_{30}$	\\
& & 3 & $D_{za1}$ & $\hat J_{50}$	\\
& & 4 & $D_{za2}$ & $\hat J_{54c}$	\\
\cline{2-5}
& \multirow{2}{*}{$B_{1g}^-$} & 1 & $\hat T_{xyz}$ & $\hat J_{32s}$	\\
& & 2 & $D_2$ & $-\hat J_{52s}$	\\
\cline{2-5}
& \multirow{2}{*}{$B_{2g}^-$} & 1 & $\hat T_{zb}$ & $\hat J_{32c}$	\\
& & 2 & $\hat D_{zb}$ & $\hat J_{52c}$	\\
\cline{2-5}
& \multirow{6}{*}{$E_{g}^-$} & 1 & $\hat J_x,\hat J_y$ & $\hat J_{11c},\hat J_{11s}$	\\
& & 2 & $\hat T_{xa},\hat T_{ya}$ & $(\sqrt{5}\hat J_{33c}-\sqrt{3}\hat J_{31c})/\sqrt{8},(-\sqrt{5}\hat J_{33s}-\sqrt{3}\hat J_{31s})/\sqrt{8}$	\\
& & 3 & $\hat T_{xb},\hat T_{yb}$ & $(\sqrt{3}\hat J_{33c}+\sqrt{5}\hat J_{31c})/\sqrt{8},(-\sqrt{3}\hat J_{33s}+\sqrt{5}\hat J_{31s})/\sqrt{8}$	\\
& & 4 & $\hat D_{xa1},\hat D_{ya1}$ & $(3\sqrt{14}\hat J_{55c}-\sqrt{70}\hat J_{53c}+2\sqrt{15}\hat J_{51c})/16,(3\sqrt{14}\hat J_{55s}+\sqrt{70}\hat J_{53s}+2\sqrt{15}\hat J_{51s})/16$	\\
& & 5 & $\hat D_{xa2},\hat D_{ya2}$ & $(\sqrt{10}\hat J_{55c}+9\sqrt{2}\hat J_{53c}+2\sqrt{21}\hat J_{51c})/16,(\sqrt{10}\hat J_{55s}-9\sqrt{2}\hat J_{53s}+2\sqrt{21}\hat J_{51s})/16$	\\
& & 6 & $\hat D_{xb},\hat D_{yb}$ & $(\sqrt{30}\hat J_{55c}+\sqrt{6}\hat J_{53c}-2\sqrt{7}\hat J_{51c})/8,(\sqrt{30}\hat J_{55s}-\sqrt{6}\hat J_{53s}-2\sqrt{7}\hat J_{51s})/8$	\\
\hline
\hline
\end{tabular}
\end{table*}

\section{Multipole operators}
Definitions of the multipole operators under $D_{4h}$ point group are listed in Table \ref{tab2} and formed by the Hermitian tensor operators (for $q\neq0$)
\begin{equation}
\begin{aligned}
& \hat J_{kqc}=\frac{1}{\sqrt{2}}[(-1)^q \hat J_{kq}+\hat J_{k,-q}],
\\
& \hat J_{kqs}=\frac{1}{\sqrt{2}i}[(-1)^q \hat J_{kq}-\hat J_{k,-q}].
\end{aligned}
\end{equation}
For $q=0$, $\hat J_{k0}$ is itself Hermitian. Different notations have been used for multipole operators in the literature \cite{Shiina1997,Kusunose2008,Kuramoto2009,Ikeda2012,Watanabe2018}. Here we follow the convention in Ref. \cite{Ikeda2012} and use the tesseral harmonics in $O_h$ point group or cubic harmonics as the basis \cite{Kusunose2008,Lage1947}. For 1D or 2D IR of $O_h$, the subscript denotes the tesseral harmonics $Z_{kq}({\bf\hat{r}})$. For 3D IR of $O_h$, the subscripts in $(\hat O_{xz},\hat O_{yz},\hat O_{xy})$ represent the basis function $(zx,yz,xy)$, while other multipoles are marked by the subscript $(x,y,z)$ with additional $a \,(b)$ denoting the $T_{1g/u}$ ($T_{2g/u}$) IR, 1 (2) for different equal rank basis in the same IR, and $g/u$ for inversion symmetric/antisymmetric. For instance, $(\hat O_{22},\hat O_{20})$ in the $E_g$ IR correspond to tesseral harmonics $r^2Z_{22}({\bf\hat{r}})\propto x^2-y^2$ and $r^2Z_{20}({\bf\hat{r}})\propto 3z^2-r^2$, respectively; $(\hat D_{xa1},\hat D_{ya1},\hat D_{za1})$ correspond to the first basis in $T_{1u}$ IR \cite{Shiina1997,Shiina2004,Kusunose2008,Ikeda2012}. For simplicity, we have dropped the label $j$ of the total angular momentum. Multipole operators belonging to different $j$-spaces may have same notations but different representation matrices. As examples, the matrices for the dipole $\hat J_z$ and octupoles $\hat T_{xa}$, $\hat T_{xyz}$ are
\begin{equation*}
\begin{aligned}
\hat J_z & = \frac{1}{2\sqrt{5}}
\begin{pmatrix}
-3 & & & \\
& -1 & & \\
& & 1 & \\
& & & 3\\
\end{pmatrix},\\
\\
\hat T_{xa} & = \frac{1}{4\sqrt{5}}
\begin{pmatrix}
& -\sqrt{3} & & 5 \\
-\sqrt{3} & & 3 & \\
& 3 & & -\sqrt{3} \\
5 & & -\sqrt{3} & \\
\end{pmatrix},\\
\\
\hat T_{xyz} & = \frac{i}{2}
\begin{pmatrix}
& & -1 & \\
& & & 1 \\
1 & & & \\
& -1 & &\\
\end{pmatrix},
\end{aligned}
\end{equation*}
in the $j=3/2$ subspace and
\begin{equation*}
\begin{aligned}
\hat J_z & = \frac{1}{\sqrt{70}}
\begin{pmatrix}
-5 & & & & & \\
& -3 & & & & \\
& & -1 & & & \\
& & & 1 & & \\
& & & & 3 & \\
& & & & & 5 \\
\end{pmatrix},\\
\\
\hat T_{xa} & = \frac{1}{12}
\begin{pmatrix}
& -3 & & \frac{5}{\sqrt{2}} & & \\
-3 & & \frac{3}{\sqrt{10}} & & 2\sqrt{5} & \\
& \frac{3}{\sqrt{10}} & & \frac{6}{\sqrt{5}} & & \frac{5}{\sqrt{2}} \\
\frac{5}{\sqrt{2}} & & \frac{6}{\sqrt{5}} & & \frac{3}{\sqrt{10}} & \\
& 2\sqrt{5} & & \frac{3}{\sqrt{10}} & & -3 \\
& & \frac{5}{\sqrt{2}} & & -3 & \\
\end{pmatrix},
\end{aligned}
\end{equation*}
\begin{equation*}
\begin{aligned}
\hat T_{xyz} & = \frac{i}{2\sqrt{6}}
\begin{pmatrix}
& & -\sqrt{5} & & & \\
& & & -1 & & \\
\sqrt{5} & & & & 1 & \\
& 1 & & & & \sqrt{5} \\
& & -1 & & & \\
& & & -\sqrt{5} & & \\
\end{pmatrix}.
\end{aligned}
\end{equation*}
in the $j=5/2$ subspace. All multipole matrices are normalized with $Q^{j\Gamma\alpha}_{lm} \rightarrow Q^{j\Gamma\alpha}_{lm} / \sqrt{\sum_{l'm'} |Q^{j\Gamma\alpha}_{l'm'}|^2}$. For  two-dimensional IR $E^{\pm}_g$, we fix the sign in the definition of its two components so that $\hat Q^{j_1\Gamma\alpha}_{r}\hat Q^{j_2\Gamma\beta}_{r} \equiv (\hat Q^{j_1\Gamma\alpha}_{x}\hat Q^{j_2\Gamma\beta}_{x}+\hat Q^{j_1\Gamma\alpha}_{y}\hat Q^{j_2\Gamma\beta}_{y})/2$ belongs to the identity representation.

\begin{figure}[b]
\begin{center}
\includegraphics[width=8.6 cm]{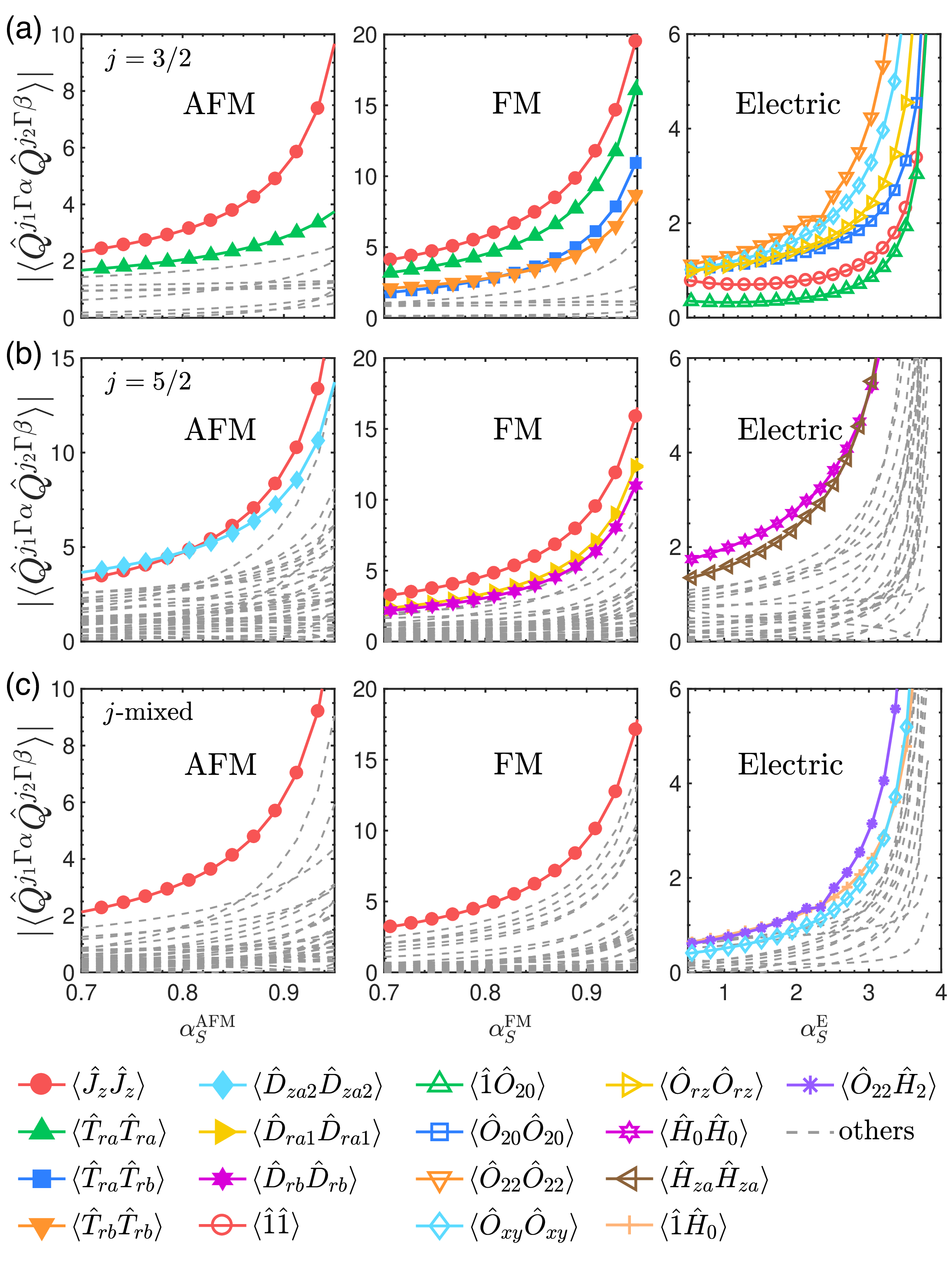}
\caption{(a) Evolution of 11 AFM and FM magnetic multipole correlations and 6 electric ones as functions of the Stoner factor $\alpha_S$ at their respective wave vector for $j=3/2$. (b)  Evolution of 38 AFM and FM magnetic multipole correlations and 23 electric ones as functions of their respective $\alpha_S$ for $j=5/2$. (c) Evolution of 30 AFM and FM magnetic multipole correlations and 15 electric ones as functions of their respective $\alpha_S$ for $j$-mixed subspace.}
\label{fig5}
\end{center}
\end{figure}

\section{Leading RPA multipole fluctuations}
We evaluate the dynamical susceptibility $\hat\chi^{\text{RPA}}$ from random phase approximation (RPA), project it into the $j$-space, and define an effective strength for each multipole channel using the correlation function \cite{Ikeda2012,Nomoto2014}, 
\begin{equation}
\langle \hat Q^{j_1\Gamma\alpha}\hat Q^{j_2\Gamma\beta}\rangle = \sum_{lml'm'} Q^{j_1\Gamma\alpha}_{ml} [\hat\chi^{\text{RPA}}]^{ll'}_{mm'}({\bf{Q}},\omega\rightarrow0) Q^{j_2\Gamma\beta}_{l'm'}.
\end{equation}
The RPA susceptibility $\hat\chi^{\text{RPA}}$ is given by
\begin{equation}
\hat \chi^{\text{RPA}}(q) = \left[ 1 - \hat\Gamma_0 \hat\chi_0(q) \right]^{-1}\hat\chi_0(q),
\end{equation}
where $q$ denotes both momentum and bosonic Matsubara frequency, and $\hat\chi_0(q)$ is the Lindhard susceptibility, 
\begin{equation}
[\chi_0]^{l_1l_2}_{l_3l_4} (q) 
= -T\sum_{k} G^0_{l_1l_2}(k) G^0_{l_4l_3}(k-q).
\end{equation}
They are calculated based on the 2D TB Hamiltonian $H_0$ with an additional local Coulomb term, 
\begin{equation}
H_{U} = \sum_{i}\sum_{l_1l_2l_3l_4} [\Gamma_0]^{l_1l_4}_{l_3l_2} c_{il_1}^\dag c_{il_2}^\dag c_{il_3} c_{il_4},
\end{equation}
where $l_i$ represents both orbital and spin quantum numbers. The interaction matrix $\hat\Gamma_0$ is given by
\begin{equation}
\begin{aligned}
\
[\Gamma_0]^{ll}_{l'l'} & = [\Gamma_0]^{l'l'}_{ll} = U ;
\\
[\Gamma_0]^{ll'}_{ll'} & = [\Gamma_0]^{l'l}_{l'l} = -U ;
\\
[\Gamma_0]^{lm}_{l'm'} & = [\Gamma_0]^{l'm'}_{lm} = J ;
\\
[\Gamma_0]^{ll'}_{mm'} & = [\Gamma_0]^{l'l}_{m'm} = -J ;
\\
[\Gamma_0]^{ll}_{m'm'} & = [\Gamma_0]^{l'l'}_{mm} = U';
\\
[\Gamma_0]^{lm'}_{lm'} & = [\Gamma_0]^{l'm}_{l'm} = -U';
\\
[\Gamma_0]^{lm}_{m'l'} & = [\Gamma_0]^{l'm'}_{ml} = J';
\\
[\Gamma_0]^{lm'}_{ml'} & = [\Gamma_0]^{l'm}_{m'l} = -J';
\\
[\Gamma_0]^{ll}_{mm} & = [\Gamma_0]^{l'l'}_{m'm'} = U'-J;
\\
[\Gamma_0]^{lm}_{lm} & = [\Gamma_0]^{l'm'}_{l'm'} = J-U',
\\
\end{aligned}
\end{equation}
where spin up or down is distinguished by the indices with or without prime. The parameters are fixed as $J=0.17U$, $J'=J$ and $U'=U-2J$ according to previous DFT+DMFT study \cite{Mravlje2011}. The obtained RPA spin susceptibility peaks at ${\bf{Q}}_{\text{RPA}}=(0.37,0.37)$, close to the experimental ${\bf{Q}}_{\text{AFM}}$. 

\begin{figure}[b]
\begin{center}
\includegraphics[width=8.6 cm]{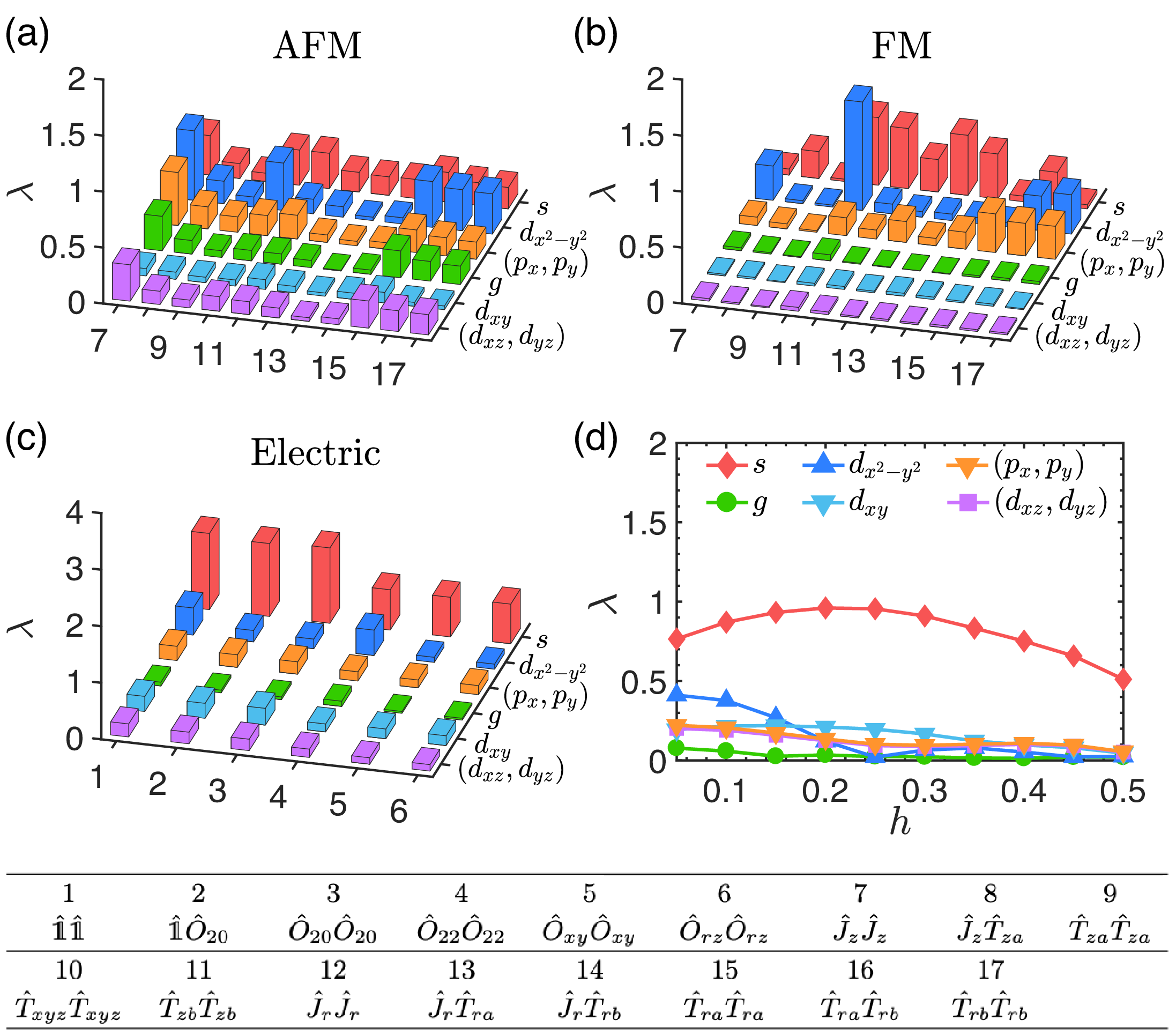}
\caption{Eigenvalues of six major pairing states, $s$ ($A_{1g}$), $d_{x^2-y^2}$ ($B_{1g}$), $(p_x,p_y)$ ($E_u$), $g$ ($A_{2g}$), $d_{xy}$ ($B_{2g}$), and $(d_{xz},d_{yz})$ ($E_g$), for individual pairing interaction in $j=3/2$ from (a) 11 AFM multipole fluctuation channels; (b) 11 FM channels; (c) 6 electric channels. (d) Eigenvalues of six major pairing states for averaged electric multipole fluctuations as a function of ${\bf{Q}}_{\text{E}}$ along $(h,h,0)$ direction. All parameters are the same as in Fig. 2 except $\xi_{z}=\xi_{xy}$.}
\label{fig6}
\end{center}
\end{figure}

\begin{figure}[b]
\begin{center}
\includegraphics[width=7.6 cm]{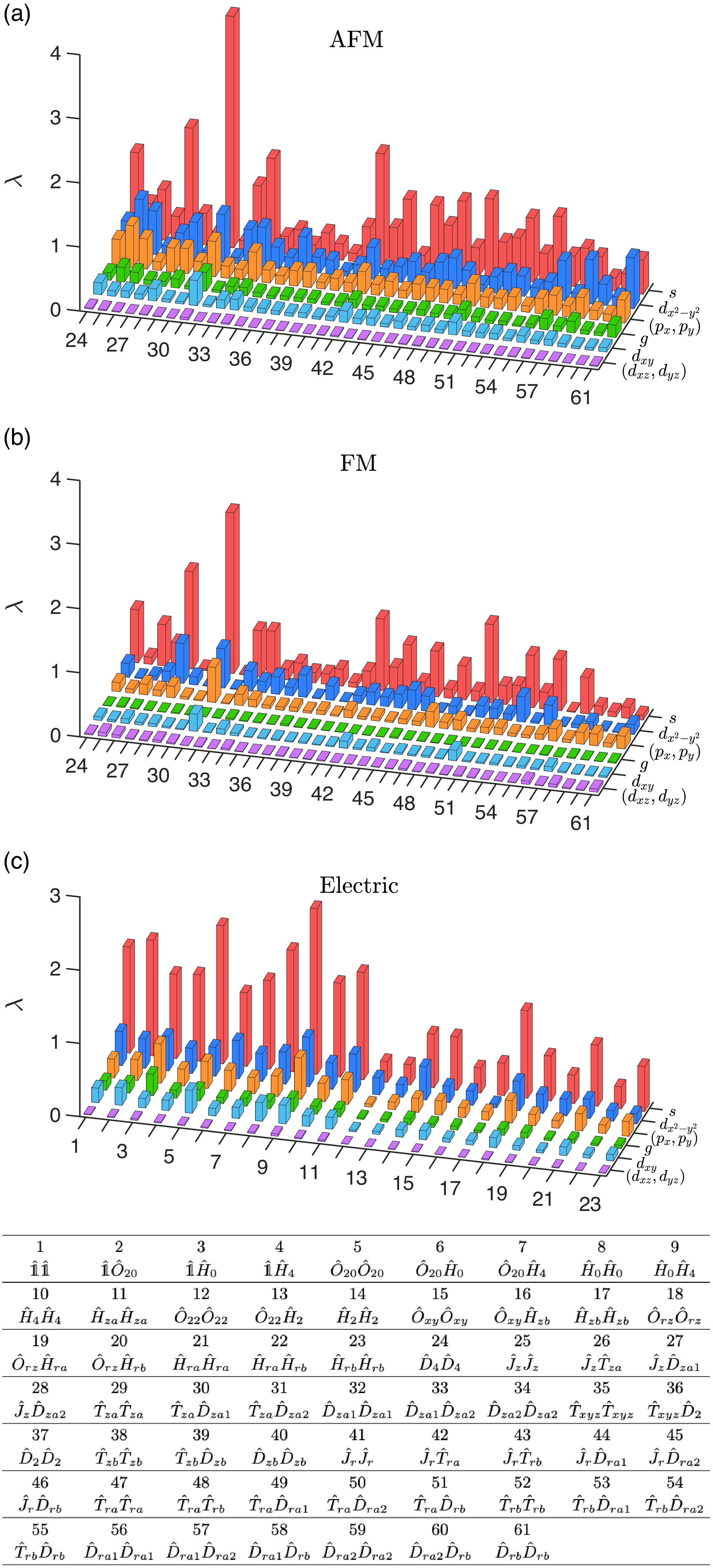}
\caption{Comparison of six major pairing states obtained by diagonalizing the linearized Eliashberg equations on the 3D Fermi surfaces of Sr$_2$RuO$_4$ from (a) 38 AFM multipole fluctuation channels; (b) 38 FM channels; and (c) 23 electric channels for $j=5/2$. All parameters are the same as in Fig. \ref{fig2}. The table on the bottom lists all multipole fluctuation channels for $j=5/2$ sorted according to their IRs and ranks.}
\label{fig7}
\end{center}
\end{figure}

Figure \ref{fig5} compares the symmetry-allowed AFM, FM, and electric multipole correlations as functions of the Stoner factor $\alpha_{S}$ at their respective wave vectors. The Stoner factors are defined as the largest eigenvalue of the matrix $\hat\Gamma_0 \hat\chi_0(q)$ for given $\textbf{Q}$ \cite{Tazai2018}. For $j=3/2$, we find, among all 11 magnetic multipoles, two leading components, $\langle \hat J_z\hat J_z\rangle$ and $\langle \hat T_{ra}\hat T_{ra}\rangle$, in the AFM channel, and four leading components, $\langle\hat J_z\hat J_z\rangle$, $\langle\hat T_{ra}\hat T_{ra}\rangle$, $\langle\hat T_{ra}\hat T_{rb}\rangle$, $\langle\hat T_{rb}\hat T_{rb}\rangle$, in the FM channel. For $j=5/2$, among all 38 magnetic multipoles, we find two leading components, $\langle\hat J_{z}\hat J_{z}\rangle$, $\langle\hat D_{za2}\hat D_{za2}\rangle$, for AFM fluctuations, and three leading components, $\langle\hat J_{z}\hat J_{z}\rangle$, $\langle\hat D_{ra1}\hat D_{ra1}\rangle$, $\langle\hat D_{rb}\hat D_{rb}\rangle$, for FM fluctuations. For $j$-mixed subspace, $\langle \hat J_z\hat J_z\rangle$ is the leading component for both AFM and FM fluctuations. The electric multipole fluctuations show no significant variation for $\alpha_S^{\text{E}}<1$ in all $j$-spaces, indicating the absence of electric instability on the RPA level.

\begin{figure}[b]
\begin{center}
\includegraphics[width=7.6 cm]{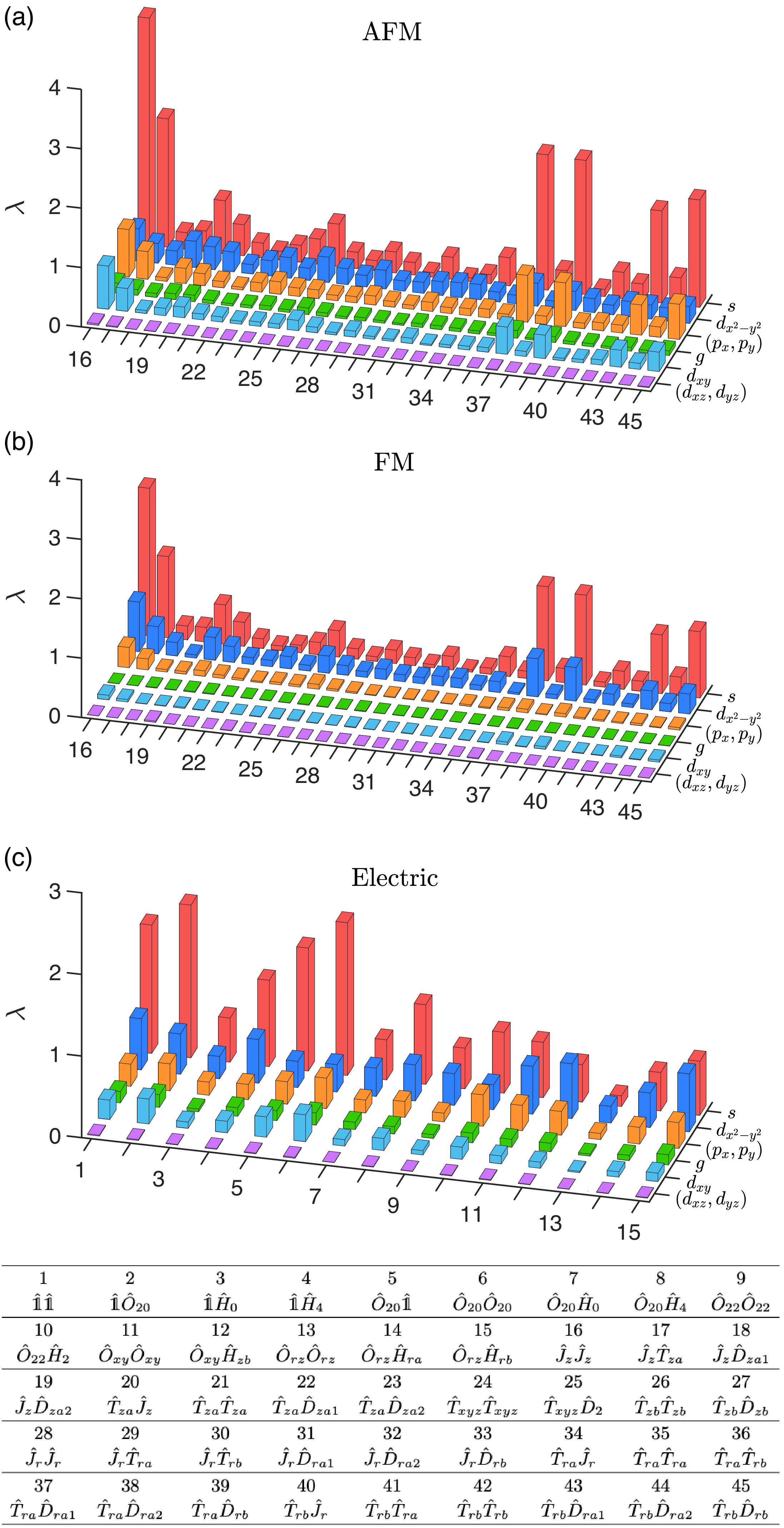}
\caption{Comparison of six major pairing states from (a) 30 AFM multipole fluctuation channels; (b) 30 FM channels; and (c) 15 electric channels in $j$-mixed subspace. All parameters are the same as in Fig. \ref{fig2}. The table on the bottom lists all multipole fluctuation channels in $j$-mixed subspace sorted according to their IRs and ranks.}
\label{fig8}
\end{center}
\end{figure}

\begin{figure}[b]
\begin{center}
\includegraphics[width=7.8 cm]{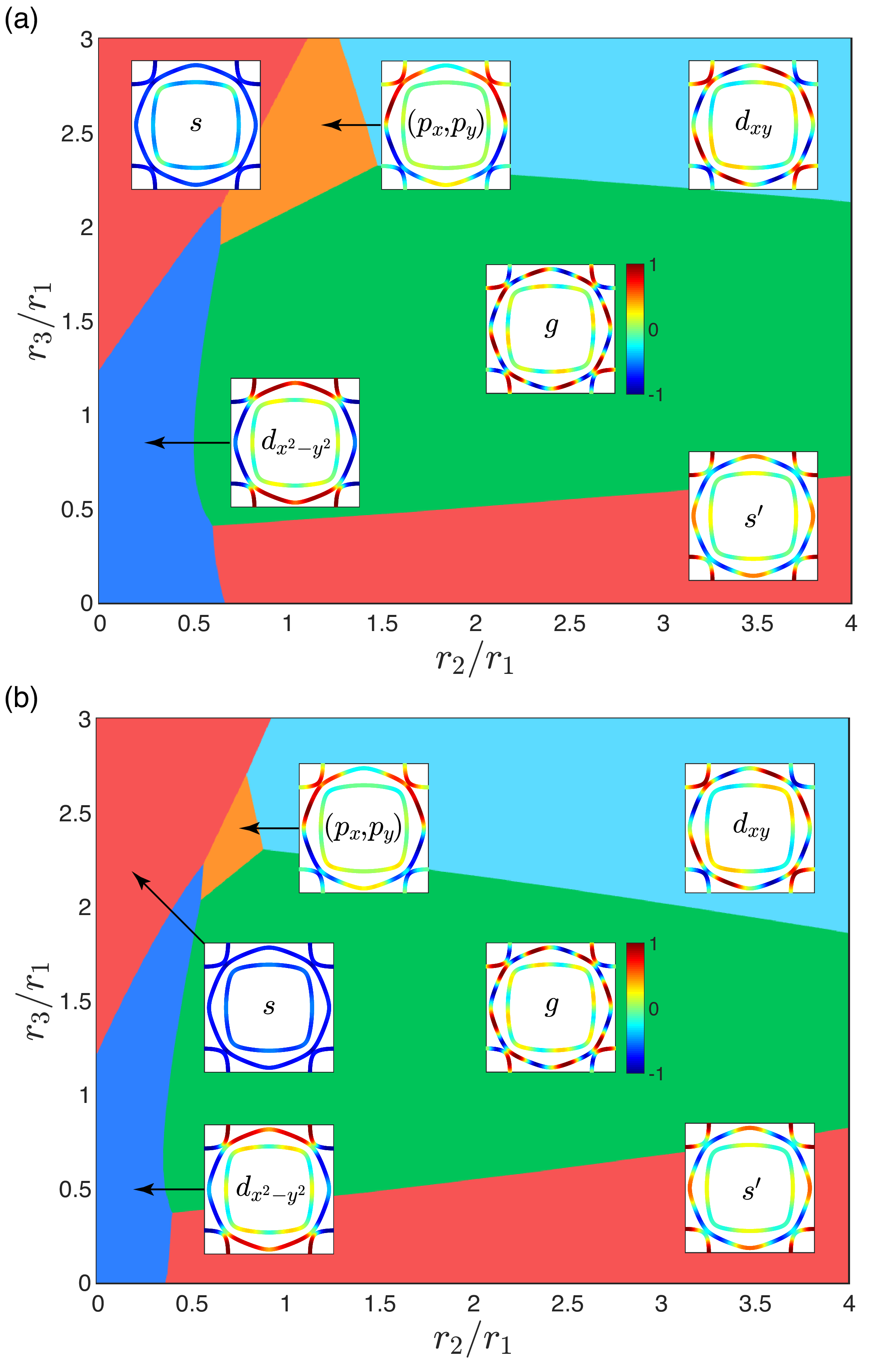}
\caption{Theoretical phase diagrams of the superconductivity in Sr$_2$RuO$_4$ for the mixed pairing interaction averaged over all multipole components in (a) $j=3/2$ and (b) $j=3/2$ and 5/2. The insets show typical gap structures in each region. All parameters are the same as in Fig. \ref{fig3}.}
\label{fig9}
\end{center}
\end{figure}

\section{Pairing states with $\xi_{z}=\xi_{xy}$ for $j=3/2$}
Figure \ref{fig6} shows six major pairing states with an artificially enlarged $\xi_{z}=\xi_{xy}$ for all three channels (AFM, FM, electric) in the $j=3/2$ manifold. All other parameters are the same as in Fig. \ref{fig2}. A larger value of $\xi_{z}$ does enhance $d_{xz}+id_{yz}$ and other $m_z$-antisymmetric pairing states, but it is still not enough to make them predominant. This implies the $d_{xz}+id_{yz}$ pairing is not favored for the quasi-2D superconductor Sr$_2$RuO$_4$, at least within our framework.

\section{Pairing states for $j=5/2$ and $j$-mixed}
Figure \ref{fig7} compares the eigenvalues of six major pairing states induced by 38 AFM or FM multipole fluctuation channels and 23 electric channels for $j=5/2$. For the FM channel, the leading dipole $\hat J_{z}\hat J_{z}$ favors $s$-wave, while the leading dotriacontapole $\hat D_{ra1}\hat D_{ra1}$ and $\hat D_{rb}\hat D_{rb}$ support ($p_x,p_y$)-wave. Figure \ref{fig8} compares the eigenvalues of six major pairing states induced by 30 AFM or FM multipole fluctuation channels and 15 electric channels for $j$-mixed subspace. Most components support $s$-wave, including the leading component $\hat J_{z}\hat J_{z}$ for AFM and FM channels and the leading components $\hat{\mathds{1}}\hat H_0$, $\hat O_{22}\hat H_2$, $\hat O_{xy}\hat O_{xy}$ for electric channel, while the rest support $d_{x^2-y^2}$. Therefore, the $j$-mixed components would enhance $s$-wave pairing in the phase diagram. We note that the off-diagonal multipole fluctuations $\hat Q^{j_1\Gamma\alpha}\hat Q^{j_2\Gamma\beta}$ with $j_1\not=j_2$ or $\alpha\not=\beta$ may lead to negative contributions to the renormalization function $Z_{\mu}$ and were usually ignored in the literature \cite{Ikeda2012,Nomoto2014,Tazai2018,Tazai2019}. They are also taken into consideration in our calculations.

\section{Robustness of $g$-wave}
Figure \ref{fig9} compares the phase diagram for the pairing interactions averaged only over $j=3/2$ and that over both $j=3/2$ and 5/2 without $j$-mixed components. The two phase diagrams are basically the same except some slight adjustment of each pairing region. While in Fig.~\ref{fig3}(b), all phase boundaries are pushed down towards smaller $r_3/r_1$ as expected once $j$-mixed contributions are included. Nevertheless, we still obtain a dominant $g$-wave in the broad intermediate region, and a two-component $d_{x^2-y^2}+ig$ pairing may emerge with strong AFM and moderate FM and electric multipole fluctuations.

\end{document}